\documentclass[12pt]{article}

\pdfoutput=1

\usepackage{graphicx}
\usepackage{dcolumn}
\usepackage{bm}
\usepackage{hyperref}
\usepackage{color}
\usepackage{amsmath}
\usepackage{amssymb}
\usepackage{amsfonts}
\usepackage{multirow}
\usepackage{cite}

\usepackage[top=30truemm,bottom=30truemm,left=25truemm,right=25truemm]{geometry}

\begin{document}


\def\a{\alpha}
\def\b{\beta}
\def\c{\varepsilon}
\def\d{\delta}
\def\e{\epsilon}
\def\f{\phi}
\def\g{\gamma}
\def\h{\theta}
\def\k{\kappa}
\def\l{\lambda}
\def\m{\mu}
\def\n{\nu}
\def\p{\psi}
\def\q{\partial}
\def\r{\rho}
\def\s{\sigma}
\def\t{\tau}
\def\u{\upsilon}
\def\v{\varphi}
\def\w{\omega}
\def\x{\xi}
\def\y{\eta}
\def\z{\zeta}
\def\D{\Delta}
\def\G{\Gamma}
\def\H{\Theta}
\def\L{\Lambda}
\def\F{\Phi}
\def\P{\Psi}
\def\S{\Sigma}

\def\o{\over}
\def\beq{\begin{align}}
\def\eeq{\end{align}}
\newcommand{\gsim}{ \mathop{}_{\textstyle \sim}^{\textstyle >} }
\newcommand{\lsim}{ \mathop{}_{\textstyle \sim}^{\textstyle <} }
\newcommand{\vev}[1]{ \left\langle {#1} \right\rangle }
\newcommand{\bra}[1]{ \langle {#1} | }
\newcommand{\ket}[1]{ | {#1} \rangle }
\newcommand{\EV}{ {\rm eV} }
\newcommand{\KEV}{ {\rm keV} }
\newcommand{\MEV}{ {\rm MeV} }
\newcommand{\GEV}{ {\rm GeV} }
\newcommand{\TEV}{ {\rm TeV} }
\newcommand{\1}{\mbox{1}\hspace{-0.25em}\mbox{l}}
\newcommand{\headline}[1]{\noindent{\bf #1}}
\def\diag{\mathop{\rm diag}\nolimits}
\def\Spin{\mathop{\rm Spin}}
\def\SO{\mathop{\rm SO}}
\def\O{\mathop{\rm O}}
\def\SU{\mathop{\rm SU}}
\def\U{\mathop{\rm U}}
\def\Sp{\mathop{\rm Sp}}
\def\SL{\mathop{\rm SL}}
\def\tr{\mathop{\rm tr}}
\def\mpl{M_{\rm Pl}}

\def\IJMP{Int.~J.~Mod.~Phys. }
\def\MPL{Mod.~Phys.~Lett. }
\def\NP{Nucl.~Phys. }
\def\PL{Phys.~Lett. }
\def\PR{Phys.~Rev. }
\def\PRL{Phys.~Rev.~Lett. }
\def\PTP{Prog.~Theor.~Phys. }
\def\ZP{Z.~Phys. }

\def\dd{\mathrm{d}}
\def\ff{\mathrm{f}}
\def\BH{{\rm BH}}
\def\inf{{\rm inf}}
\def\ev{{\rm evap}}
\def\eq{{\rm eq}}
\def\SM{{\rm sm}}
\def\Mpl{M_{\rm Pl}}
\def\GeV{{\rm GeV}}

\def\Msusy{m_{\rm stop}}
\newcommand{\Red}[1]{\textcolor{red}{#1}}
\newcommand{\TL}[1]{\textcolor{blue}{\bf TL: #1}}

\baselineskip 0.7cm
\begin{titlepage}

\vskip 1.35cm
\begin{center}
{\bf\large Supersymmetric D-term Twin Higgs}

\vskip 1.2cm
Marcin Badziak$^{1,2,3}$ and Keisuke Harigaya$^{2,3}$
\vskip 0.4cm
$^1${\it Institute of Theoretical Physics, Faculty of Physics, University of Warsaw, ul.~Pasteura 5, PL--02--093 Warsaw, Poland}\\
$^2${\it Department of Physics, University of California, Berkeley, California 94720, USA}\\
$^3${\it Theoretical Physics Group,  Lawrence Berkeley National Laboratory, Berkeley, California 94720, USA}
\vskip 1.5cm

\abstract{
We propose a new type of supersymmetric Twin Higgs model where the $SU(4)$ invariant quartic term is provided by a $D$-term potential of a new $U(1)$
gauge symmetry. In the model the 125 GeV Higgs mass can be obtained for stop masses below 1 TeV, and a tuning required to obtain the correct
electroweak scale can be as low as 20\%. A stop mass of about 2 TeV is also possible with tuning of order $\mathcal{O}(10)$~\%.
}

\end{center}
\end{titlepage}

\setcounter{page}{2}

\section{Introduction}

The main two pieces of information obtained with the Large Hadron Collider (LHC) so far is the discovery of the Standard Model (SM)-like Higgs boson
with a mass of about 125 GeV,
and no signs of New Physics close to the electroweak (EW) scale which put strong lower bounds on masses of new particles. The bounds are
especially stringent for new colored states, for which they vary between several hundreds of GeV up to about 2 TeV. These bounds threaten many
extensions of the SM that aim to solve the hierarchy problem, since naturalness requires that  the top quark contribution to the quadratic
divergence of the Higgs mass squared
is approximately cancelled by the corresponding contribution from top quark partners. If the top quark partners are heavier than the top quark
fine-tuning is reintroduced. This is known as the little hierarchy problem.

An interesting solution to the little hierarchy problem is provided by Twin Higgs models~\cite{Chacko:2005pe,Chacko:2005vw,Chacko:2005un,Falkowski:2006qq,Chang:2006ra,Batra:2008jy} which recently gained renewed interests
\cite{Craig:2013fga,Geller:2014kta,Barbieri:2015lqa,Low:2015nqa,Cheng:2015buv,Csaki:2015gfd,Contino:2017moj,Barbieri:2016zxn,Chacko:2016hvu,Craig:2016lyx,Garcia:2015loa,Garcia:2015toa,Craig:2015xla,Farina:2015uea,Freytsis:2016dgf,Farina:2016ndq,Prilepina:2016rlq,Craig:2015pha,Beauchesne:2015lva,Craig:2016kue,Yu:2016bku,Yu:2016swa,Katz:2016wtw,Cheng:2016uqk}. In this class of models the SM-like Higgs is a
pseudo-Nambu-Goldstone boson of a global $SU(4)$ symmetry, and the $\mathbb{Z}_2$ symmetry relating the SM with a mirror (or twin) SM eliminates
the quantum correction to the Higgs mass squared from the explicit breaking of the $SU(4)$ symmetry.
A key feature of this scenario is that the top quark partners are not charged under the SM color gauge group and easily evade accelerator bounds.

It should
be emphasized that Twin Higgs models do not solve the hierarchy problem but only postpone the scale at which new particles
charged under the SM color gauge group enter. Therefore, these models require  some UV completions.
Twin Higgs models have been embedded in supersymmetric (SUSY) \cite{Falkowski:2006qq,Chang:2006ra,Craig:2013fga,Katz:2016wtw} and composite Higgs
\cite{Batra:2008jy,Geller:2014kta,Barbieri:2015lqa,Low:2015nqa,Cheng:2015buv,Csaki:2015gfd,Cheng:2016uqk,Contino:2017moj} models. In the present work we
focus on SUSY UV completions.

A successful SUSY Twin Higgs model should possess at least two features. First: a large $SU(4)$ invariant Higgs quartic term
$\lambda$ to suppress the quadratic corrections to the Higgs mass parameter. More precisely, the tuning of a given model is relaxed by a factor
$2\lambda/\lambda_{\rm SM}$, as compared to the corresponding model without the mirror symmetry, where $\lambda_{\rm SM}\approx0.13$ is the SM Higgs
quartic coupling. Second: The Higgs mass of 125 GeV is obtained for
stop masses that do not lead to excessive tuning, say no worse than $\mathcal{O}(10)$~\%. In the limit of arbitrary large $\lambda$ the second
requirement would be automatically satisfied (see eq.~(\ref{eq:higgsmass_tree})). However, in realistic models there is some upper bound on $\lambda$ which does not allow tuning to go away completely.
Therefore, when discussing tuning of a given model both features should be taken into account. 

Another important point is that in phenomenologically viable  Twin
Higgs models (SUSY or not) the  $\mathbb{Z}_2$ symmetry must be broken. This is because the 125 GeV Higgs couplings measured at the LHC are close to the
SM prediction \cite{Higgscomb} and set a lower bound on the vacuum expectation value (vev) of the mirror Higgs. This results in an irreducible
tuning of $\mathcal{O}(10\mathchar`-50)$~\%,%
\footnote{This irreducible tuning may be evaded by introducing hard $\mathbb{Z}_2$ breaking but explicit models of this type require total tuning of
$\mathcal{O}(10)$~\% anyway~\cite{Katz:2016wtw}.}
depending on the amount of the Higgs invisible decays to mirror particles and other details of a given model.%
\footnote{Cosmological constraints on Twin Higgs models generically require non-negligible Higgs decays to mirror fermions \cite{Barbieri:2016zxn}.
See, however, refs.~\cite{Chacko:2016hvu,Craig:2016lyx}.   For other studies of cosmological implications of Twin Higgs models see
e.g.~refs.~\cite{Garcia:2015loa,Garcia:2015toa,Craig:2015xla,Farina:2015uea,Freytsis:2016dgf,Farina:2016ndq,Prilepina:2016rlq}.}
 On the
other hand, $\mathbb{Z}_2$ breaking is beneficial as far as the Higgs mass is concerned because in the limit of maximal $\mathbb{Z}_2$ breaking the
tree-level Higgs mass is enhanced by a factor $\sqrt{2}$ with respect to the prediction of the Minimal Supersymmetric Standard Model (MSSM). 
This makes SUSY Twin Higgs models also attractive for relatively light stops - satisfying the current experimental constraints but within the
ultimate reach of the LHC.
One of the
goals of the
present paper is to quantify the gain in the Higgs mass and study implications for the stop masses paying particular attention to effects of
$SU(4)$ and $\mathbb{Z}_2$ breaking. In particular, we determine parameter space in which tuning does not
exceed the irreducible tuning from the Higgs coupling measurements discussed above and calculate upper bounds on stop masses under this assumption.

We find that existing SUSY Twin Higgs models cannot saturate the irreducible tuning.
In models proposed so far the $SU(4)$ invariant quartic term  is generated by an $F$-term of a singlet chiral field
\cite{Falkowski:2006qq,Chang:2006ra,Craig:2013fga,Katz:2016wtw}. The $SU(4)$ invariant quartic term
is then maximized for $\tan\beta=1$ and decreases as $\sin^2(2\beta)\approx 4/\tan^2\beta$.
On the other hand, the $SU(4)$ breaking quartic coupling from the EW $D$-term, which contributes to the Higgs mass, is an increasing function of $\tan
\beta$,
and hence a smaller $\tan \beta$ requires a larger stop mass.%
\footnote{Suppression of the Higgs mass at small $\tan\beta$ can be avoided if $\mathbb{Z}_2$ breaking quartic term is present but this comes at a
cost of model simplicity, see e.g.~refs.~\cite{Falkowski:2006qq,Katz:2016wtw}.}
As a result the 125 GeV Higgs mass is incompatible with a large $SU(4)$ invariant quartic term and sufficiently light stops that do not lead to large fine-tuning.
We also find that the higgsino mass is required to be small to suppress the singlet-Higgs mixing, which would otherwise reduce the Higgs mass. 

Motivated by these findings we propose a new type of supersymmetric Twin Higgs model where the $SU(4)$ invariant quartic term is provided by a
$D$-term
potential of a new $U(1)_X$ gauge symmetry. In this setup the $SU(4)$ invariant quartic term grows with $\tan\beta$, which does not conflict with the Higgs
mass constraint. We discuss the Landau pole constraints  and show that the $SU(4)$ invariant quartic term can be large enough to minimize the tuning
in
the regime where the model is under perturbative control.
We present scenarios in which the tuning of the EW scale is solely determined by the irreducible one while the LHC constraints on sparticle masses are satisfied. In the least tuned region stops are within the reach of the LHC.
Even if no sparticles are found at the end of the high-luminosity run of the LHC the tuning of the model may be still better than 10~\%.

The rest of the paper is organized as follows. In section~\ref{sec:THreview} we briefly review the $F$-term Twin Higgs model, introduce the
$D$-term model and discuss constraints from perturbativity. In section~\ref{sec:decoupling} we discuss the impact of the Higgs mass on SUSY Twin
Higgs models in a quite general effective field theory framework assuming that the only source of the tree-level $SU(4)$ breaking quartic term is the
EW $D$-term potential. In section~\ref{sec:FtermDterm} we discuss the fine-tuning of SUSY Twin Higgs models in detail. 
We show that the non-decoupling effect of the singlet have a substantial impact on the Higgs mass, which worsens fine-tuning in the $F$-term model,
while analogous effects are almost absent in the $D$-term model. We quantify the naturalness of the $D$-term model in several scenarios. We briefly discuss differences in the heavy Higgs spectrum and phenomenology between $F$-term and $D$-term models. We reserve section~\ref{sec:concl} for
our concluding remarks.

\section{SUSY Twin Higgs models}
\label{sec:THreview}

In this section we briefly review a SUSY Twin Higgs model in which an $SU(4)$ invariant quartic term is generated via an $F$-term potential and introduce a
new class of SUSY Twin Higgs models in which an $SU(4)$ invariant quartic term is generated via a $D$-term potential. 

\subsection{$F$-term Twin Higgs}

A SUSY realisation of the Twin Higgs mechanism was first proposed in refs.~\cite{Falkowski:2006qq,Chang:2006ra} which used an $F$-term of a singlet chiral superfield $S$ to
generate the $SU(4)$ invariant quartic term. The $F$-term Twin Higgs model was analysed in light of the Higgs boson discovery in ref.~\cite{Craig:2013fga}, and more
recently in ref.~\cite{Katz:2016wtw}. The $SU(4)$ invariant part of the $F$-term model is given by the following superpotential and soft SUSY breaking
terms:
\begin{align}
\label{V_Fterm}
 &W_{ SU(4)}= (\mu+\lambda_S S)(H_u H_d +H'_uH'_d) +  \mu' S^2 \,, \\
 &V_{ SU(4)}= m_{H_u}^2(|H_u|^2 + {|H'_u|}^2) + m_{H_d}^2(|H_d|^2 + {|H'_d|}^2) - b (H_u H_d +H'_uH'_d + {\rm h.c.}) + m_S^2 |S|^2 \,.
\end{align}
Note that the $SU(4)$ symmetry is automatically realised by the $\mathbb{Z}_2$ symmetry.
At tree level, the $SU(4)$ symmetry is explicitly broken by the EW $D$-term potential:
\begin{equation}
\label{eq:EWDterm}
 V_D=\frac{g^2+g'^2}{8} \left[ (|H_u|^2-|H_d|^2)^2 + (|H'_u|^2-|H'_d|^2)^2 \right] \,.
\end{equation}
The above terms are $\mathbb{Z}_2$ invariant. In phenomenologically viable models the $\mathbb{Z}_2$ symmetry must be broken. This is obtained by
introducing soft scalar masses:
\begin{equation}
 V_{\rm soft} = \Delta m_{H_u}^2 H_u^2 + \Delta m_{H_d}^2 H_d^2 + \Delta b (H_u H_d  + {\rm h.c.}) \,.
\end{equation}
The Twin Higgs mechanism may relax fine-tuning only if the $SU(4)$ invariant quartic term $\lambda$ is larger than the SM Higgs quartic coupling. In this
model this coupling is given, after integrating out a heavy singlet and heavy Higgs bosons, by
\begin{equation}
 \lambda=\lambda_S^2\frac{\sin^2\left( 2\beta \right)}{4}\equiv \lambda_F.
\end{equation}
So large $\lambda$ prefers large $\lambda_S$ and small $\tan\beta$. However, there is an upper bound on $\lambda_S$ and a lower bound on $\tan\beta$.
The former constraint comes from the requirement of perturbativity. Avoiding a Landau pole below 10 (100) times the singlet mass scale requires
$\lambda_S$ below about 1.9 (1.4). A lower bound on $\tan\beta$ originates from the Higgs mass constraint which we discuss in more detail in the
following sections.

\subsection{$D$-term Twin Higgs}

As an alternative to the $F$-term Twin Higgs model we propose a model in which a large $SU(4)$ invariant quartic term originates from a
non-decouping $D$-term of a new $U(1)_X$ gauge symmetry. Such a non-decoupling $D$-term may be present if the mass of a scalar field responsible for
the breaking of the $U(1)_X$ gauge symmetry is dominated by a SUSY breaking soft mass, see Appendix for details. Such models were considered in the
context of non-twinned SUSY in
refs.~\cite{Langacker:1999hs,Batra:2003nj,Morrissey:2005uz,Cheung:2012zq,Endo:2011gy,Huo:2012tw,DAgnolo:2012vzj,Craig:2012bs,Bertuzzo:2014sma,
Capdevilla:2015qwa}. The non-decoupling $D$-term potential can be written as
\begin{equation}
\label{eq:VU1X}
 V_{U(1)_X}=\frac{g_X^2}{8} \left( |H_u|^2-|H_d|^2 + |H'_u|^2-|H'_d|^2 \right)^2 \left(1-\epsilon^2\right) \,,
\end{equation}
where $\epsilon$ is a model-dependent parameter in the range between 0 and 1. We refer to the Appendix for explicit model that naturally allows for
$\epsilon\ll1$ which maximizes the magnitude of the $D$-term potential.
This term gives the following $SU(4)$ invariant coupling:
\begin{equation}
 \lambda=g_X^2\frac{\cos^2\left( 2\beta \right)}{8}\left(1-\epsilon^2\right) \equiv \lambda_D \,.
\end{equation}
A crucial difference with the $F$-term model is that $\lambda$ is now maximized in the limit of large $\tan\beta$ which makes it easier to satisfy
the Higgs mass constraint. This merit of a $D$-term generated $SU(4)$ invariant quartic term was recently noted also in ref.~\cite{Katz:2016wtw}. The
magnitude of $\lambda$ is still bounded from above to avoid too low a Landau pole scale so it is not guaranteed that fine-tuning is considerably
relaxed. 

The beta function of the 
$U(1)_X$ gauge coupling constant depends on the charge assignment of particles in the visible and mirror sectors.
Let us first assume that the $U(1)_X$ charges of the MSSM
particles and the mirror particles are a linear combination of $U(1)_Y$ and $U(1)_{B-L}$ charges, so that the gauge anomaly is cancelled solely by
introducing the right-handed neutrinos,
\begin{align}
q_X = q_Y + x q_{\rm B-L} \,.
\end{align}
Then the beta function of the $U(1)_X$ gauge coupling constant is given by
\begin{align}
\frac{\rm d}{{\rm dln} \mu}\frac{8\pi^2}{g_X^2} = b_X,\nonumber\\
b_X = - (32 x^2 + 32 x+22).
\end{align}
The scale of the Landau pole is maximized when $x=-1/2$, which we assume in the following.
In this case, $b_X = -14$. For fraternal Twin Higgs models~\cite{Craig:2015pha}, where the mirror of the first and the second generations are not
introduced, $b_X = -10$.

Denoting the mass of the $U(1)_X$ gauge boson as $m_X$,  the scale of the Landau pole $M_{c}$ is given by
\begin{align}
M_c = m_X \times {\rm exp} [ - \frac{8\pi^2}{g_X(m_X)^2 b_X }].
\end{align}
We expect that the Twin Higgs theory has a UV completion at the scale $M_c$.%
\footnote{Since all the SM fermions are charged under the $U(1)_X$ symmetry, they are expected to be described as a (partially) composite particles around the scale $M_c$.}
We require that $M_c$ is larger than the mediation scale of the SUSY breaking which we assume throughout the article to be $\Lambda=100 \Msusy$,
where $\Msusy$ is the soft mass of stops. In order to avoid the experimental constraints on $m_X$, to be discussed later, the mass of $X$ is typically
expected to be a factor of between 5 to 10 larger than the stop masses. This requires $M_c \gtrsim 10m_X$ which sets an upper bound on $g_X(m_X)$ of
about 1.6 (1.9) for the mirror (fraternal) Twin Higgs model.

The constraint is relaxed if the $U(1)_X$ charge is flavor dependent.
For example, it is possible that the first and the second generation fermions are $U(1)_X$ neutral, and their yukawa couplings are generated via
mixing
between these fermions and heavy $U(1)_X$ charged fermions.
Then the renormalization group (RG) running of the $U(1)_X$ gauge coupling constant is significant only above the masses of those heavy fermions, and
below those mass scales
$b_X=
-6$, which allows values of $g_X(m_X)$ up to about 2.4 if one requires $M_c \gtrsim 10m_X$. 
In this type of models, the experimental lower bound on $m_X$ which is discussed later is also significantly relaxed.
Throughout this paper we refer
to this class of models as flavor
non-universal SUSY $D$-term Twin Higgs models. 
Such a construction is also motivated by the observed hierarchy of fermions masses and explains why the SM fermions of the third generation 
are much heavier than those of the first two generations. Nevertheless, to also explain the observed hierarchy among the first two generations of
the SM fermions ala Froggatt-Nielsen~\cite{Froggatt:1978nt}, additional horizontal symmetry would be required, see
e.g.~refs.~\cite{Cohen:1996vb,Dine:1993np,Pomarol:1995xc,Barbieri:1995uv,Dudas:1995eq,Dudas:1996fe} for the ideas of SUSY model building in this direction and its relation to possible solutions of
the SUSY flavor problem.

\section{SUSY Twin Higgs in decoupling limit}
\label{sec:decoupling}

Before going to a disscussion of full SUSY Twin Higgs models it is instructive to discuss general effective theory with heavy MSSM-like Higgs
doublets and other
states decoupled. In such a case the Higgs potential depends only on the SM-like Higgs and its mirror partner:
\begin{align}
V = \lambda (|H'|^2 + |H|^2)^2 -m^2 (|H'|^2 + |H|^2) + \Delta \lambda(|H'|^4 + |H|^4) + \Delta m^2 |H^2| \,.
\end{align}
The first two terms are both $\mathbb{Z}_2$ and $SU(4)$ symmetric, $\Delta \lambda$ preserves $\mathbb{Z}_2$ but breaks $SU(4)$, while $\Delta
m^2$ breaks both $\mathbb{Z}_2$ and $SU(4)$ symmetry. One could also consider a hard $\mathbb{Z}_2$ breaking quartic term which in our setup is
subdominant, see ref.~\cite{Katz:2016wtw} for discussion of effects of hard $\mathbb{Z}_2$ breaking.
The vevs of the Higgs fields and the masses of them are given by
\begin{align}
v'^2=\vev{H'}^2 = \frac{m^2}{4 \lambda} \frac{ 1 + \frac{\lambda \Delta m^2}{\Delta \lambda m^2}}{1 + 2 \Delta \lambda/\lambda},~~
v^2 = \vev{H}^2 = \frac{m^2}{4 \lambda} \frac{ 1 - \frac{\lambda \Delta m^2}{\Delta \lambda m^2} - \frac{\Delta m^2} {m^2}}{1 + 2 \Delta \lambda/\lambda},
\end{align}
\begin{align}
m_h^2 =&
2\left( \lambda + \Delta \lambda \right) \left( v'^2 + v^2  \right) 
- 2\sqrt{ \left( \lambda + \Delta \lambda \right)^2 \left( v'^2 + v^2  \right)^2 - 4 \Delta \lambda \left( 2 \lambda + \Delta\lambda \right) v'^2 v^2 },\\
m_{h'}^2 =&
2\left( \lambda + \Delta \lambda \right) \left( v'^2 + v^2  \right) 
+ 2\sqrt{ \left( \lambda + \Delta \lambda \right)^2 \left( v'^2 + v^2  \right)^2 - 4 \Delta \lambda \left( 2 \lambda + \Delta\lambda \right) v'^2 v^2
}.
\end{align}

The above formulae are independent of whether the UV completion is supersymmetric or not. In SUSY models the $SU(4)$ symmetry is generically broken at
tree level by the EW D-term potential of eq.~\eqref{eq:EWDterm} which in the above framework corresponds to
\begin{align}
 \Delta \lambda  \supset \frac{g^2+g'^2}{8} \cos^2 \left( 2\beta \right) \equiv \Delta \lambda_{\rm SUSY} \approx 0.07  \cos^2 \left( 2\beta \right).
\end{align}
Note that $\Delta \lambda_{\rm
SUSY}$ grows as a function of $\tan\beta$ from zero (for $\tan\beta=1$) up to $0.07$ in the large $\tan\beta$ limit.
Thus for lower $\tan\beta$ the observed Higgs mass gives a stronger lower bound
on masses of stops which dominate the radiative corrections to the Higgs mass.

Let us first discuss the Higgs mass at the tree level.
In the limit of an exact $\mathbb{Z}_2$ symmetry and a large $SU(4)$
preserving quartic coupling, $\lambda \gg \Delta \lambda$, the tree-level Higgs mass is the same as in MSSM. However, in phenomenologically viable
models the $\mathbb{Z}_2$ symmetry must be broken. Moreover, corrections to the Higgs mass of order $\mathcal{O}(\Delta \lambda / \lambda)$ are often
non-negligible in realistic SUSY Twin Higgs models. After taking these effects into account the tree-level Higgs mass in SUSY Twin Higgs models is
approximately given by
\begin{equation}
\label{eq:higgsmass_tree}
 \left(m_h^2\right)_{\rm tree} \approx 2 M_Z^2 \cos^2 \left( 2\beta \right) \left(1-\frac{v^2}{f^2} \right) + \mathcal{O}(\Delta \lambda
/ \lambda)  \,,
\end{equation}
where the first term is the effect of $\mathbb{Z}_2$ breaking while the second term corresponds to the correction of order $\mathcal{O}(\Delta \lambda
/
\lambda)$, which is negative, and $f^2 \equiv v^2 + v^{'2}$. We see that in the limit $v\ll f$ and $\lambda \gg \Delta \lambda$ the tree-level Higgs
mass is enhanced by a factor of $\sqrt{2}$ with
respect to the MSSM Higgs mass which in large $\tan\beta$ limit turns out to be very close to the observed Higgs mass of 125 GeV. This is another
virtue of SUSY Twin Higgs
models. While  large hierarchy between $v$ and $f$, which introduces the fine-tuning of
\begin{align}
\Delta_{v/f} = \frac{1}{2} \left( \frac{f^2}{v^2} -2\right),
\end{align}
is not preferred from the view point of naturalness, the ratio $f/v$ above about two or three, which is required
by the Higgs coupling measurements, leads to a significant boost of the tree-level Higgs mass. Terms $\mathcal{O}(\Delta \lambda / \lambda)$ also
reduce the Higgs mass and in the limit of $\lambda \ll \Delta \lambda$, in which the $SU(4)$ symmetry is not even approximately realized, the Higgs mass
is the same as that in the MSSM.

Due to the large value of the top Yukawa couplings the quantum correction by the top and the stop significantly affect the Higgs mass.
We take into account the quantum correction by computing the Coleman-Weinberg (CW) potential of the Higgs fields.
We include contributions from top
and stop from both visible and mirror sectors. In reliable prediction of the Higgs mass proper choice of a renormalization scale for the top Yukawa
coupling $y_t$ is crucial since the correction to the Higgs mass is proportional to $y_t^4$. It is well known that in MSSM the Higgs mass calculated
at one loop level grossly overestimates the full result if $y_t$ at the top mass scale is used, see e.g.~refs.~\cite{Carena:1995bx,Carena:1995wu,HHH}.
In
ref.~\cite{HHH} it was shown that the dominant two-loop effects in the computation of the Higgs mass can be accommodated by using in the one loop result
the RG running top mass at a scale $\mu_t\equiv\sqrt{m_t \Msusy}$. Since the RG running at one loop in the visible and mirror sector is independent
from each other, we expect that using in the CW potential $y_t$ matched to the top mass at a scale $\mu_t$ will also accommodate the leading
two-loop corrections. Therefore, in our calculations we adopt the RG-improved procedure of ref.~\cite{HHH} using their formulae with $m_t(m_t)=165$
GeV. 

Since we do not include corrections other than that from top/stop loops, some non-negligible theoretical uncertainties may still be present even after the RG
improvement. We estimate this uncertainty by comparing our result in the limit $f=v/\sqrt{2}$ and $\lambda \gg \Delta \lambda$, in which the MSSM
Higgs mass should be recovered, with {\tt SOFTSUSY} \cite{softsusy} computation of the MSSM Higgs mass for a degenerate sparticle spectrum (but with
heavy MSSM-like Higgs decoupled) and find that our procedure still overestimates the Higgs mass by about 5 (3) GeV for the stop masses of 1 TeV (400
GeV). These numbers are in good agreement with findings of ref.~\cite{HHH}. In Twin Higgs model this overestimation may be even larger expecially for
$\Msusy \gg f$, because in such a case also the mirror stop contributes substantially to the Higgs mass. On top of that, there are additional
contributions to the Higgs mass arising from mass splitings in sparticle spectrum, which are unavoidable given strong LHC bounds on the gluino mass, that
typically result in further reduction of the Higgs mass in MSSM. On the other hand, the Higgs mass may be enhanced by few GeV by stop mixing effects
(not included in our computation) with only a minor increase in tuning caused by the stop sector.\footnote{For maximal stop mixing the Higgs mass may
be enhanced by as much as 10 GeV but that would increase EW tuning by a factor of about four
(for a given value of stop soft masses) so we do not consider maximal stop mixing as optimal choice from the naturalness perspective.}
Having all of the above in mind we substract 5 GeV from the Higgs mass obtained using the above procedure and assume theoretical uncertainty of 3
GeV.

\begin{figure}[t]
\centering
\includegraphics[width=.49\textwidth]{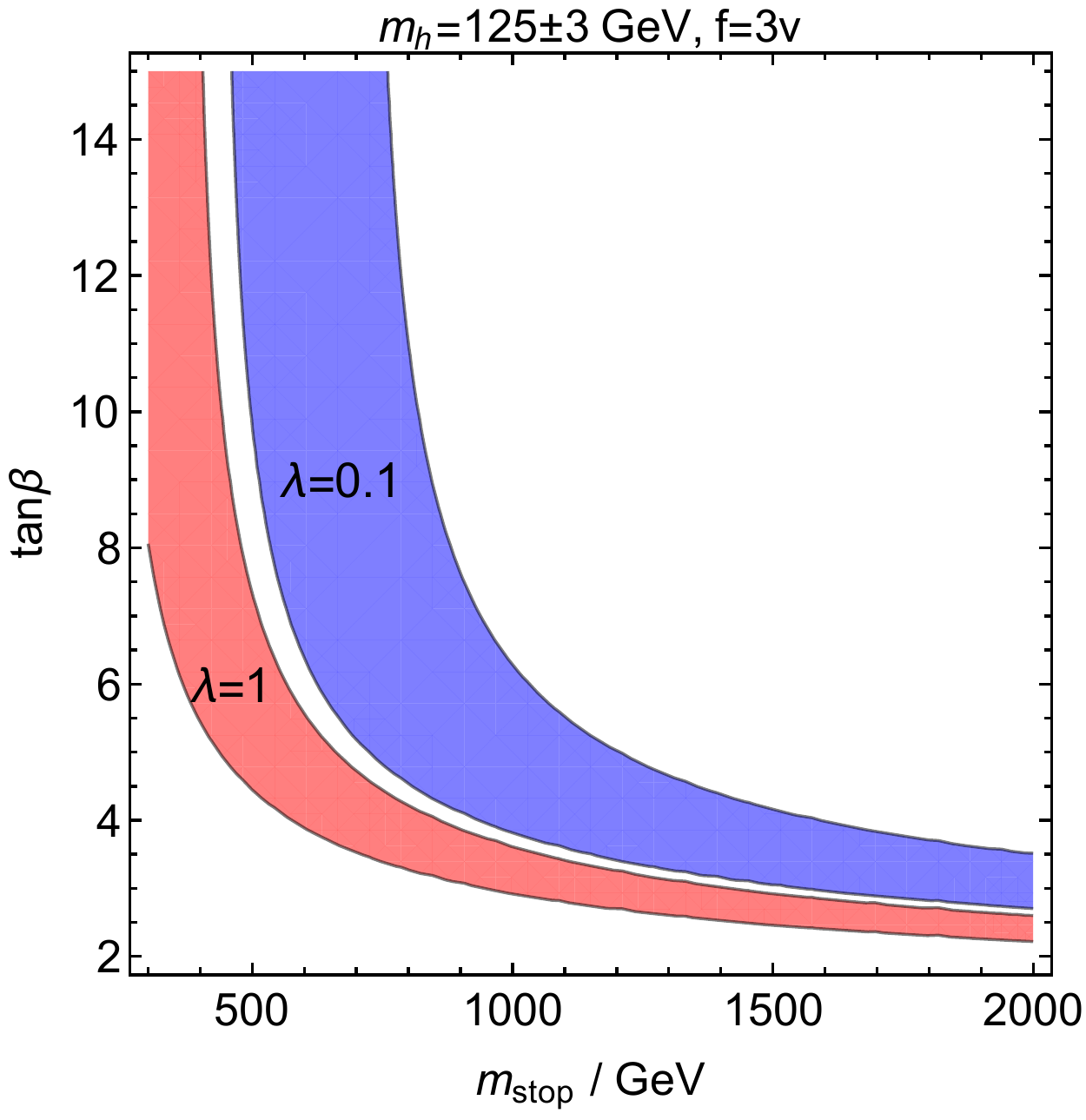}
\includegraphics[width=.49\textwidth]{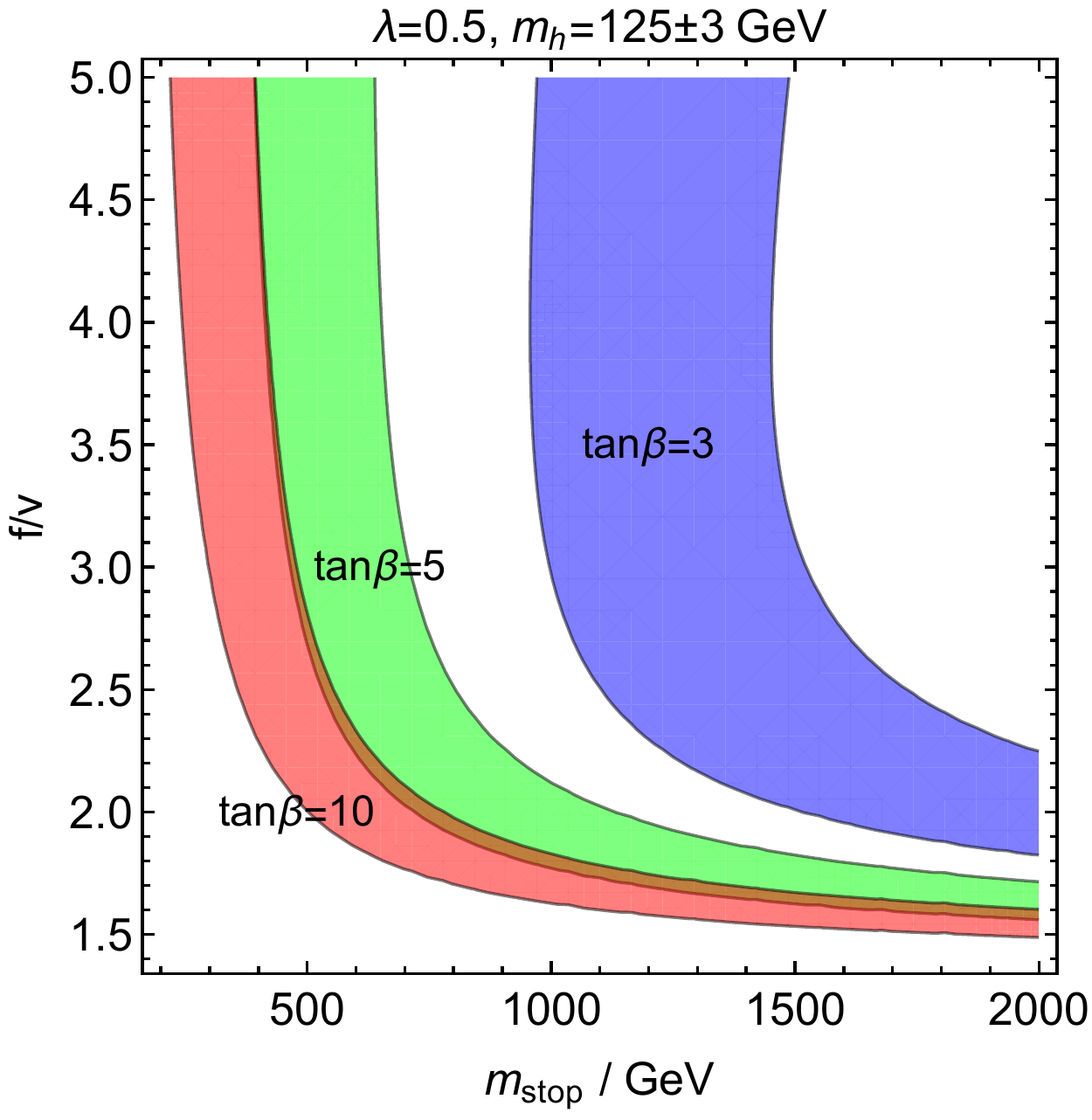}
\caption{The SM-like Higgs mass of $m_h=125\pm3$ GeV in the plane $\Msusy$-$\tan\beta$ for fixed $f=3v$ with $\lambda=0.1$ and $\lambda=1$ (left
panel)
and in the plane $\Msusy$-$f/v$ for $\tan\beta=3,\ 5,\ 10$ (right panel) assuming the decoupling limit of SUSY Twin Higgs model.
}
\label{fig:higgsmass}
\end{figure}

In the left panel of fig.~\ref{fig:higgsmass} the region preferred by the measured Higgs mass is presented in the plane $\Msusy$-$\tan\beta$.
It is clear from this plot that much lighter stops are sufficient to satisfy the Higgs mass constraint than in the MSSM even for a $SU(4)$
preserving quartic coupling of similar size as the one from the $SU(4)$ breaking EW D-term. In particular, a lower bound on $\tan\beta$ is much weaker
but it should be emphasized that values of $\tan\beta\lesssim 3$ cannot accommodate the measured Higgs mass for sub-TeV stops even for large $\lambda$.
The preferred range of stop masses does not depend strongly on $f/v$ as long as it is above about 2.5, i.e. in a region preferred by the Higgs
coupling measurements, as seen from the right panel of fig.~\ref{fig:higgsmass}. 

The Higgs mass larger than the MSSM one also results in a rather strong upper bound on stop masses for large $\tan\beta$. In fact in the limit of large
$\tan\beta$ and large $\lambda$ for $f=3v$ the stops must be lighter than about 400 GeV, as seen in
fig.~\ref{fig:higgsmass}. While 400~GeV stops may be still consistent with the LHC constraints if the LSP mass is heavier than about 300~GeV~\cite{LHCstop}, the 400 GeV left-handed sbottom (which has a similar mass to the left-handed stop) is already excluded by the LHC~\cite{LHCsbottom}.
The
upper bound on the stop/sbottom masses may be relaxed to about 600 GeV for $f=2.3v$ -- the smallest value of $f$ consistent with the data
\cite{Katz:2016wtw}, which may evade the current constraints if the LSP mass is above about 500 GeV. The upper bound may be further relaxed if
non-decoupling effects of the remaining scalars are important but one should keep in
mind that generically the LHC has the ability to set an upper bound on $\tan\beta$. 

\begin{figure}[t]
\centering
\includegraphics[width=.49\textwidth]{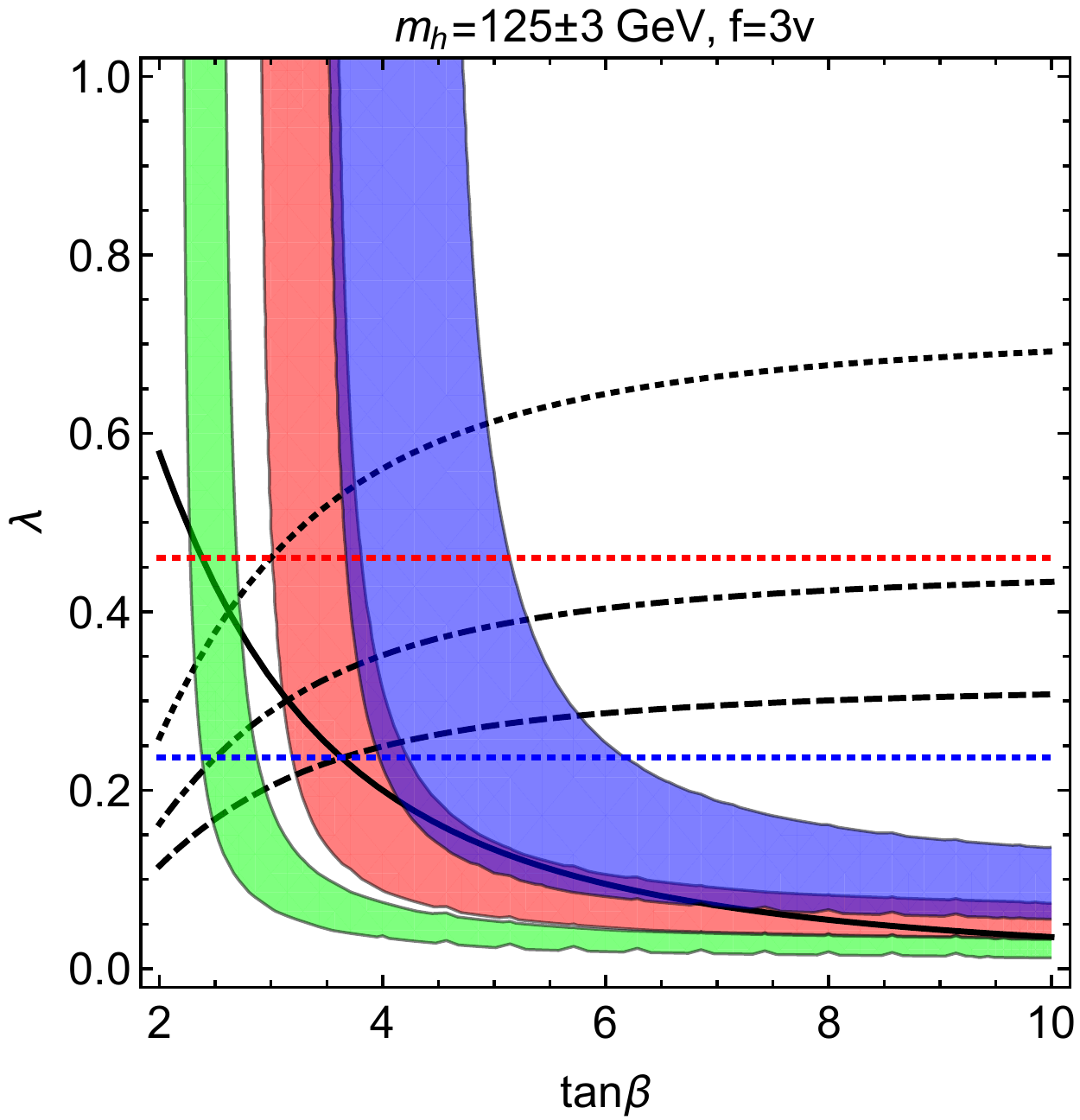}
\includegraphics[width=.49\textwidth]{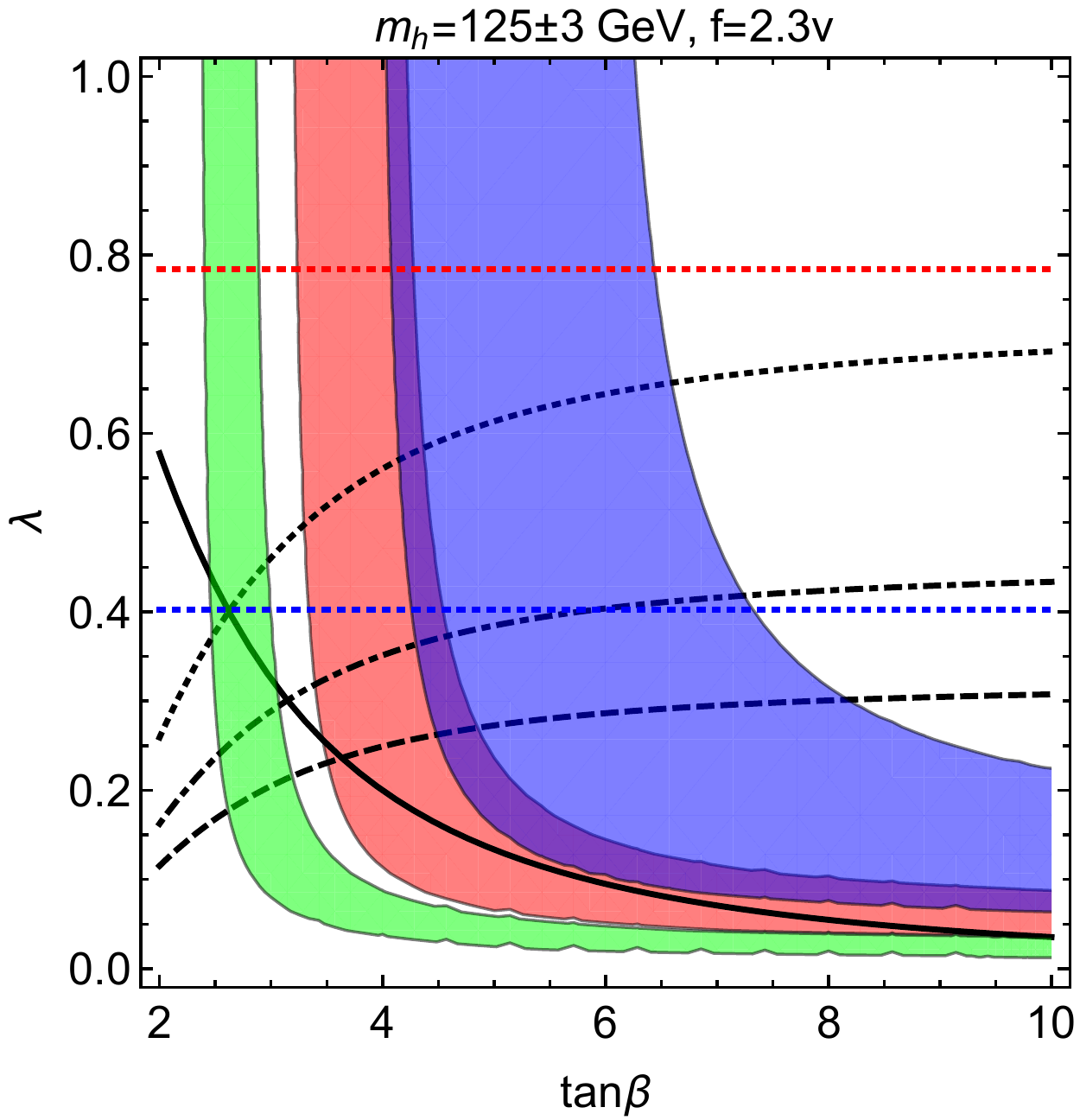}
\caption{The SM-like Higgs mass of $m_h=125\pm3$ GeV in the plane $\tan\beta$-$\lambda$ for several values of the stop masses: 700 GeV (blue), 1
TeV
(red) and 2 TeV (green) in the decoupling limit of SUSY Twin Higgs models. Horizontal dotted lines correspond to minimal value of $\lambda$ for which
there is
no tuning from the stop sector assuming the cut-off scale $\Lambda=100\Msusy$. The black solid lines show $\lambda_F$ with $\lambda_S=1.9$ i.e.~the value for which the Landau pole is ten times above the singlet mass scale in the $F$-term model. The dashed, dash-dotted, dotted lines show
$\lambda_D$ with $\epsilon=0$ and $g_X=1.6,\ 1.9,\ 2.4$ i.e.~the value for which the Landau pole is ten times above the $X$ gauge boson mass scale for the 
mirror, fraternal and flavor non-universal $D$-term Twin Higgs model, respectively. In the left (right) panel $f/v=3$ ($2.3$).
}
\label{fig:higgsmass2}
\end{figure}

In SUSY UV completions one generally expects that $\lambda$ depends on $\tan\beta$. This is the case in models where the $SU(4)$ invariant quartic term
is generated from $F$-term as well as in the case of $D$-term generated $\lambda$ that we
propose in
the present paper. In fig.~\ref{fig:higgsmass2} we plot the Higgs mass in the plane $\tan\beta$-$\lambda$ for several values
of the stop masses. As expected from the previous discussion, the lighter stops are the larger $\tan\beta$ and $\lambda$ are preferred by the Higgs
mass
constraint. In fig.~\ref{fig:higgsmass2} we also present maximal values of $\lambda$ as a function of $\tan\beta$ in the $F$-term and $D$-term Twin
Higgs models under assumption that the Landau pole scale is at least ten times larger than the singlet mass (the $X$ gauge boson mass) in the
$F$-term ($D$-term) models.

We see that for $\Msusy$ up to 1 TeV the maximal value of $\lambda$ is definitely larger in the $D$-term Twin Higgs models than in the $F$-term one,
especially in the flavor non-universal version of the former, so the improvement in tuning as compared to non-twinned SUSY models is better in the
$D$-term model. Larger
$\lambda$ in the $F$-term model than the $D$-term one may be obtained for $\Msusy=2$ TeV but is not large enough to prevent tuning from the stop
sector which in the leading-log approximation given by
\begin{equation}
 \Delta_{f,{\rm stop}}^{\rm LL}\approx \frac{3y_t^2}{8\pi^2\lambda f^2}\Msusy^2\ln\left(\frac{\Lambda}{\Msusy}\right) \,,
\end{equation}
where $\Lambda$ is the messenger scale that we take to be $100\Msusy$.

Fig.~\ref{fig:higgsmass2} emphasizes that in the $F$-term model it is hard to saturate the irreducible tuning
even in the
decoupling limit. Moreover, this situation gets much worse after taking into account non-decoupling effects, as we discuss in the next section. On the
other hand, the $D$-term model can saturate this tuning even if $f/v$ is as small as 2.3 which corresponds to $\Delta_{v/f}\approx1.6$ so essentially
no tuning exists at all. In the $D$-term case non-decoupling effects are much less important than in the $F$-term one and the total tuning can be
$\mathcal{O}(10)$~\% as we show in the next section.

\section{$F$-term vs $D$-term Twin Higgs beyond decoupling limit}
\label{sec:FtermDterm}

In this section we give a more detailed analysis of $F$-term and $D$-term Twin Higgs models, going beyond the decoupling limit.
We quantify the degree of fine-tuning by introducing the measure,
\begin{align}
\Delta_v \equiv &  \Delta_f \times \Delta_{v/f},\nonumber  \\
\Delta_f = & {\rm max}_i |\frac{\partial{\rm ln} f^2}{\partial{\rm ln} x_i(\Lambda)}| .
\end{align}
Here $x_i(\Lambda)$ are the parameters of the theory evaluated at the mediation scale of the SUSY breaking.
To evaluate $\Delta_f$ we solve the renormalization group equations (RGEs) of parameters between $\Msusy$ and $\Lambda$ at the one-loop level.

\subsection{$F$-term Twin Higgs model} 
\begin{figure}[t]
\centering
\includegraphics[width=.49\textwidth]{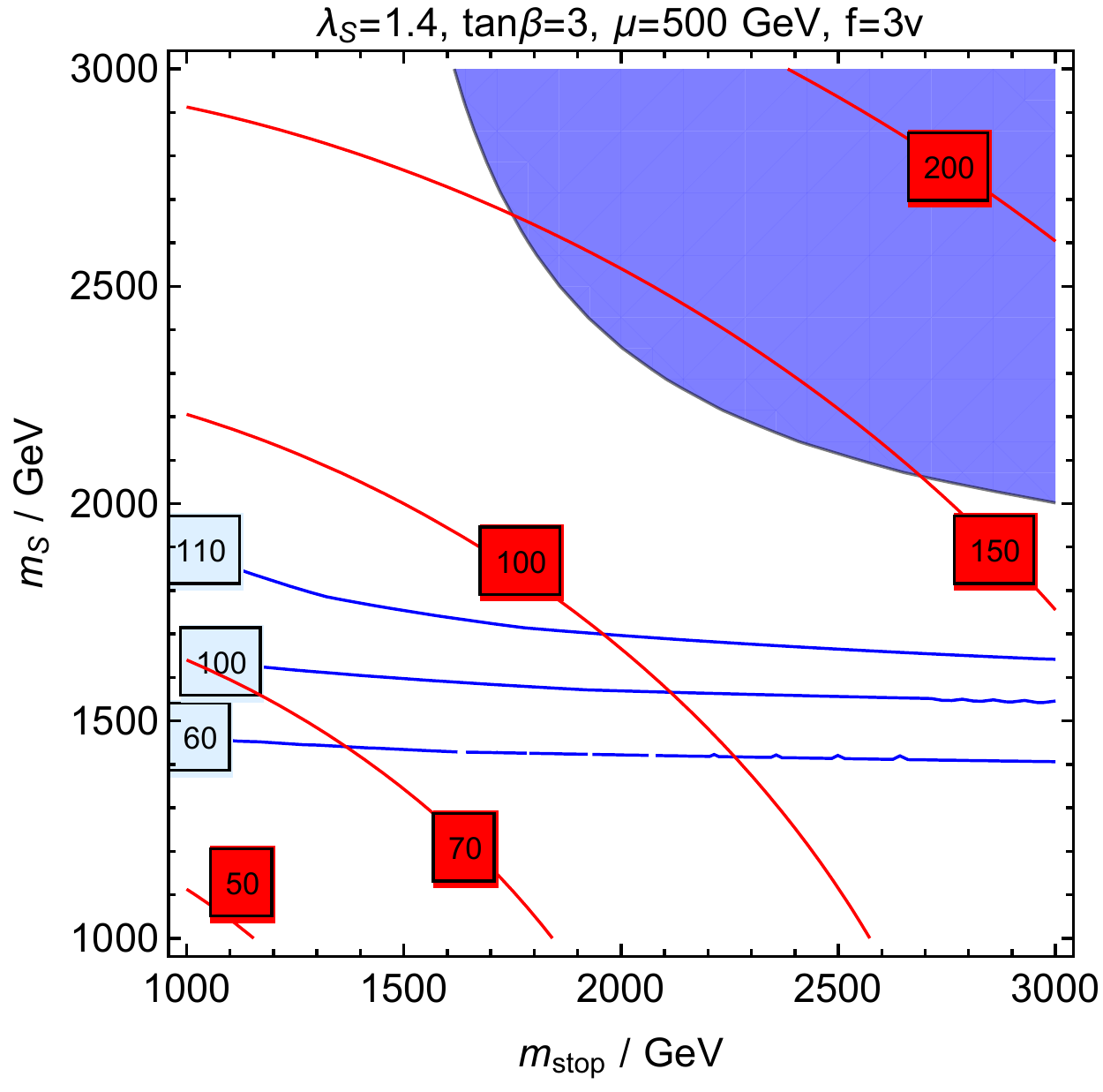}
\includegraphics[width=.49\textwidth]{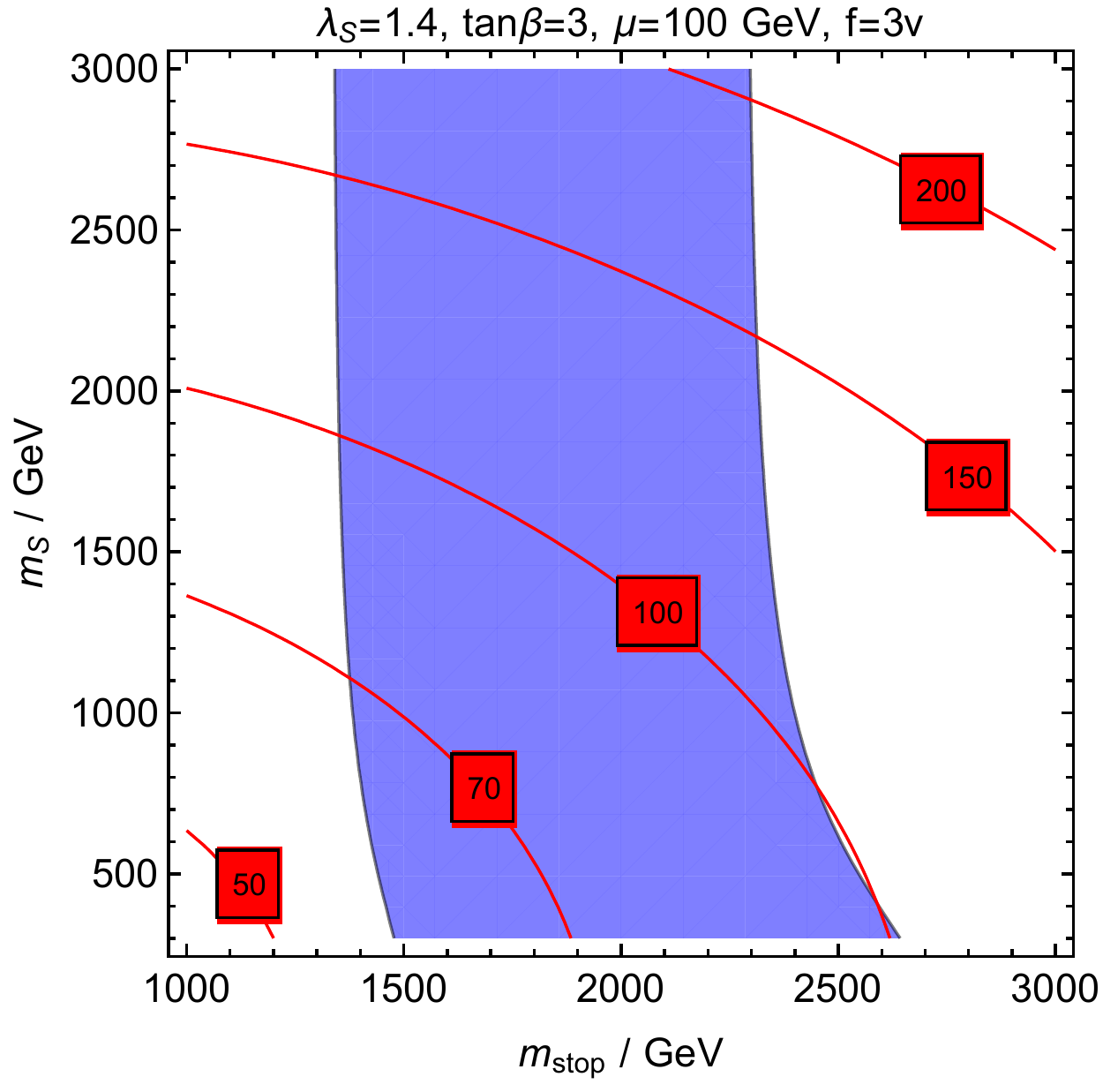}
\caption{The SM-like Higgs mass of $m_h=125\pm3$ GeV (blue region) in the plane $m_{\rm stop}$-$m_S$ for $\lambda_S=1.4$, $f=3v$,   $\tan\beta=3$,
$m_A=1.5$ TeV, $M_3=2$ TeV and $\mu'=0$ in the $F$-term Twin Higgs model. In the left
(right) panel $\mu=500$ (100) GeV. The red contours correspond to total fine-tuning of the model. The blue contours in the left panel depict values
of $m_h$.}
\label{fig:mh_mS_mstop}
\end{figure}

It was already noted in ref.~\cite{Craig:2013fga} that fine-tuning is not minimized for the maximal value of $\lambda_S$ that avoids the Landau pole
because in
such a case the tuning from large soft singlet mass dominates that from stops. Instead, the fine-tuning is minimized for some intermediate value of
$\lambda_S$ in the range between 1 and 1.5 which results in $\Delta_v\sim 50\div100$, i.e. $1\div2\%$ fine-tuning \cite{Craig:2013fga}. This result
for
fine-tuning in the $F$-term Twin Higgs model was obtained for $\mu=500$ GeV, $m_S=1\ {\rm TeV}$ and $m_{\rm stop}=2$ TeV and was confirmed recently in
ref.~\cite{Katz:2016wtw}. However, we find that the fine-tuning in the $F$-term Twin Higgs model is even more severe due to the Higgs mixing with the
singlet which gives a negative contribution to the Higgs mass. The Higgs-singlet mixing is proportional to $\lambda_S v \mu$.
For large $\lambda_S$ (which is crucial in the Twin Higgs mechanism) and moderate values of $\mu$ (which naturally is close to $f$) the mixing is sizable and cannot be neglected in the
Higgs mass calculation for the singlet mass of 1 TeV. This is demonstrated in fig.~\ref{fig:mh_mS_mstop} from which it is clear that the correct
Higgs mass requires, for $\mu=500$ GeV, the singlet mass of at least $2$ TeV, while for $m_S=1$ TeV the Higgs direction turns out to be tachyonic. However,
for values of $m_S$ above $2$ TeV the fine-tuning from the heavy singlet dominates the one from
stops. In consequence, the fine-tuning is worse than 1\%. The problem is less severe for smaller values of $\mu$ which, however, is
constrained from below, $\mu\gtrsim 100$ GeV by null results of the LEP chargino searches~\cite{Heister:2002mn}. It can be seen from the right panel
of fig.~\ref{fig:mh_mS_mstop} that for $\mu=100$ GeV there is small impact of the 1 TeV singlet
on the Higgs mass and the fine-tuning can still be at the level of $1\div2\%$. Note that for $\mu=100$ GeV, $m_S$ can be very small without making
the Higgs direction tachyonic because for small $m_S$ the physical singlet mass is dominated by that given by the mirror Higgs vev, $\lambda_S f$. 
\footnote{The presence of the singlet also modifies the Higgs couplings but in the regions of parameter space consistent with the observed Higgs mass
the singlet component of the SM-like Higgs is at most few percent so the constraints from the Higgs coupling measurements are easily satisfied. The
$H_u$ and $H_d$ components of the singlet-like state are also at most few percent 
so there are no meaningful constraints from direct LHC searches for additional scalars. }
The Higgs-singlet mixing may be reduced if a non-vanishing singlet $A$-term
of the form $A_\lambda\lambda H_u H_d S$ is introduced. In such a case this mixing can be suppressed for any value of $\mu$ if $A_\lambda\approx \mu
\tan\beta$. However, if $\mu$ is not small this implies large $A_\lambda$ which again generates large fine-tuning. We conclude that in the $F$-term
Twin Higgs model fine-tuning at least at the level of few percent  is required, and a very light higgsino is a signature of the smallest fine-tuning,
similarly as in the MSSM, in
spite of the Twin Higgs mechanism.

\subsection{$D$-term Twin Higgs model}

In the $D$-term Twin Higgs model there are no effects that significantly affect the prediction for the Higgs mass in the decoupling limit analysed in section \ref{sec:decoupling} so the Higgs mass is determined by the value of $\lambda$, $\tan\beta$ and $f/v$. Let us know discuss fine-tuning in
this model and show that it is significantly better than the one in the $F$-term Twin Higgs model. Apart from the usual tuning from stops, the tuning may also
arise from a threshold correction to the soft Higgs mass which is proportional to
a new gauge boson mass squared:
\begin{equation}
\label{deltamHu_X}
 \left(\delta m_{H_u}^2\right)_{X}= \frac{g_X^2}{64 \pi^2} m_X^2 \ln\left(\epsilon^{-2}\right) \,.
\end{equation}
It is important to note that, in contrast to the $F$-term model, this correction does not depend on the cut-off scale. However, it does depend on a 
model-dependent parameter $\epsilon$ which characterizes the size of the mass splitting in the vector supermultiplet. The same parameter enters the
effective $SU(4)$-preserving quartic coupling:
\begin{equation}
\label{eq:lambdaD}
 \lambda= \frac{g_X^2}{8}\cos^2(2\beta) (1-\epsilon^2) \,. \nonumber
\end{equation}
Therefore, small values of $\epsilon$ are preferred to maximize the $SU(4)$-preserving quartic term but this enhances the threshold correction to
the soft Higgs mass of eq.~\eqref{deltamHu_X}. There is a lower bound on the size of this correction which comes from searches for additional $U(1)$
gauge bosons. For large values of $g_X$ the most stringent constraint comes from searches for off-shell production of the $X$ boson in dimuon final states
at LEP which gives a lower bound of $m_X\gtrsim 4350$ GeV $\times g_X$~\cite{Cheung:2012zq}.\footnote{The LHC constraints on $m_X$ are becoming
competitive with the LEP one, especially for smaller values of $g_X$. However, we found that for $g_X\gtrsim1$ the recent LHC constraints
\cite{Atlas_Xlimits} are still weaker than the LEP one.  } Since the limit is stronger for larger $g_X$ the fine-tuning is not necessarily smaller for
larger $g_X$.

\begin{figure}[p]
\centering
\includegraphics[width=.49\textwidth]{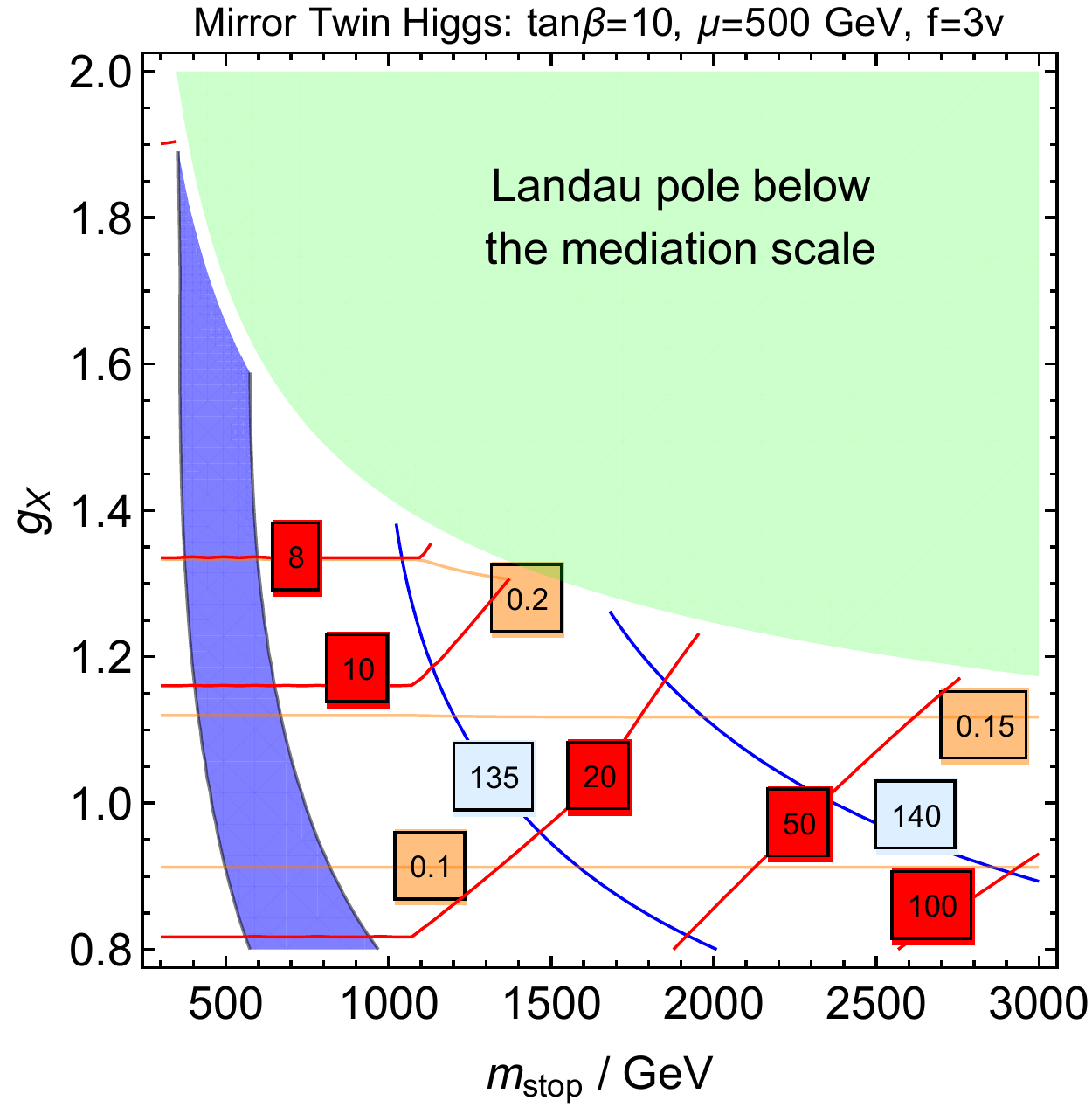}
\includegraphics[width=.49\textwidth]{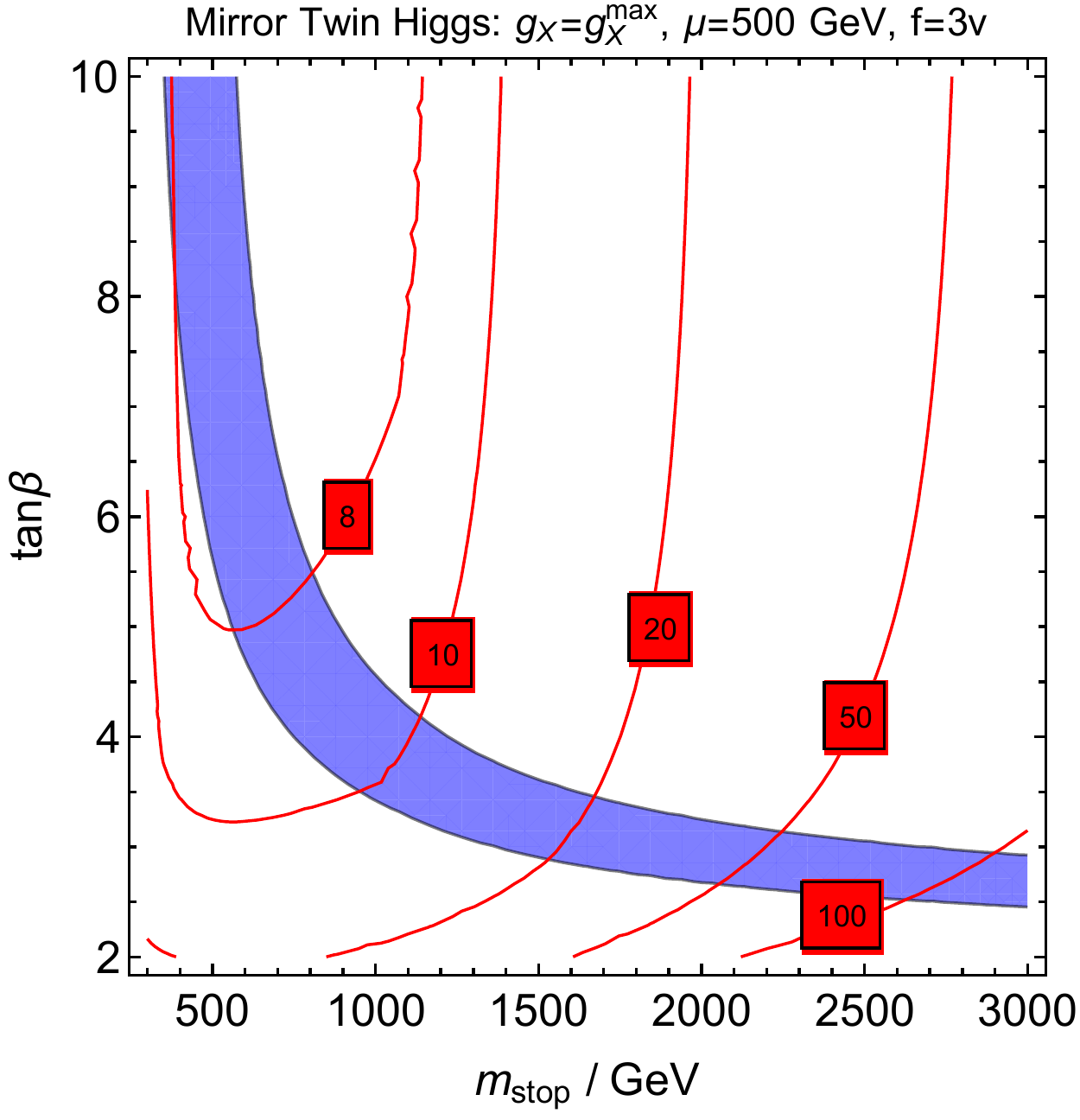}
\includegraphics[width=.49\textwidth]{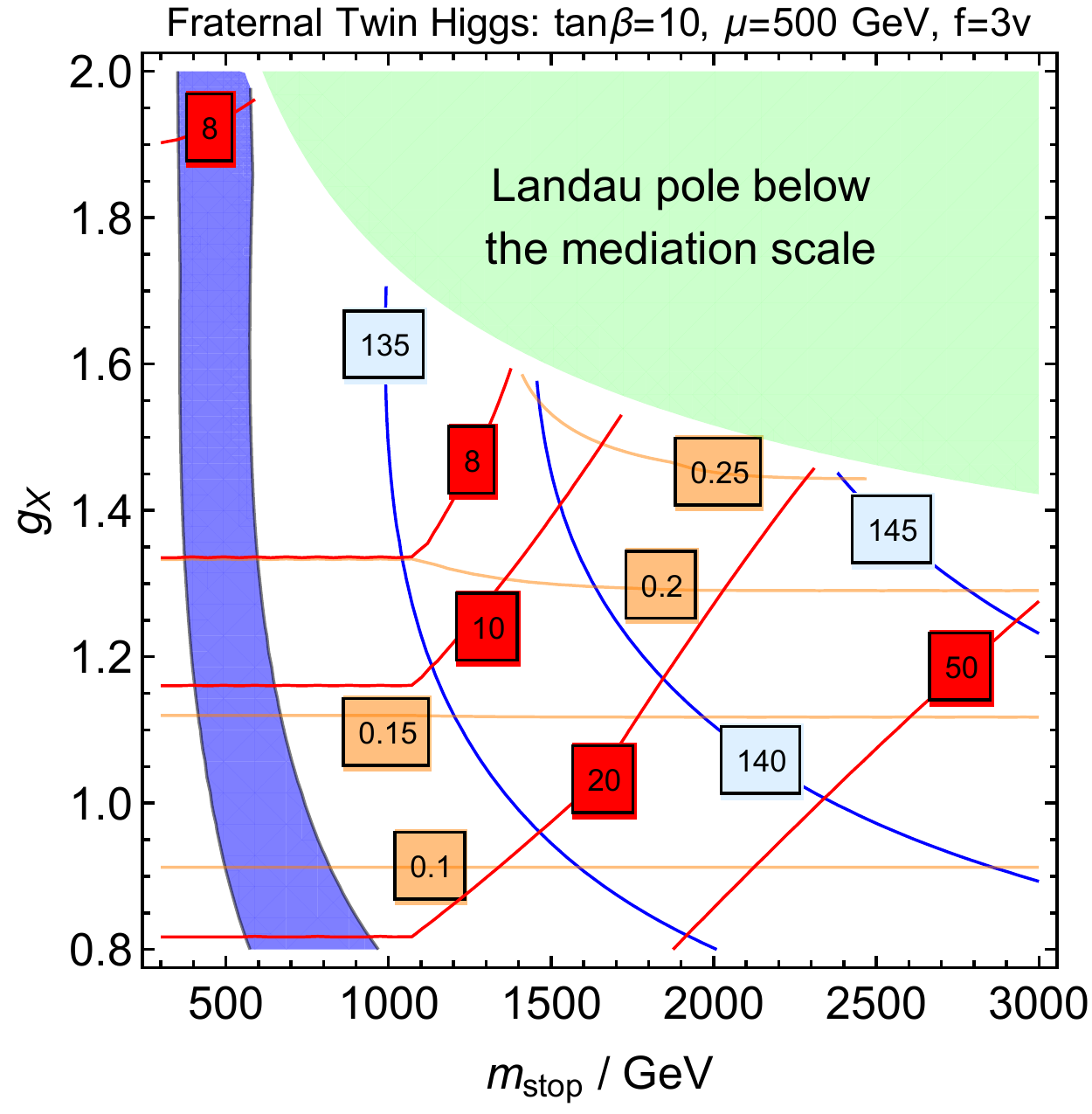}
\includegraphics[width=.49\textwidth]{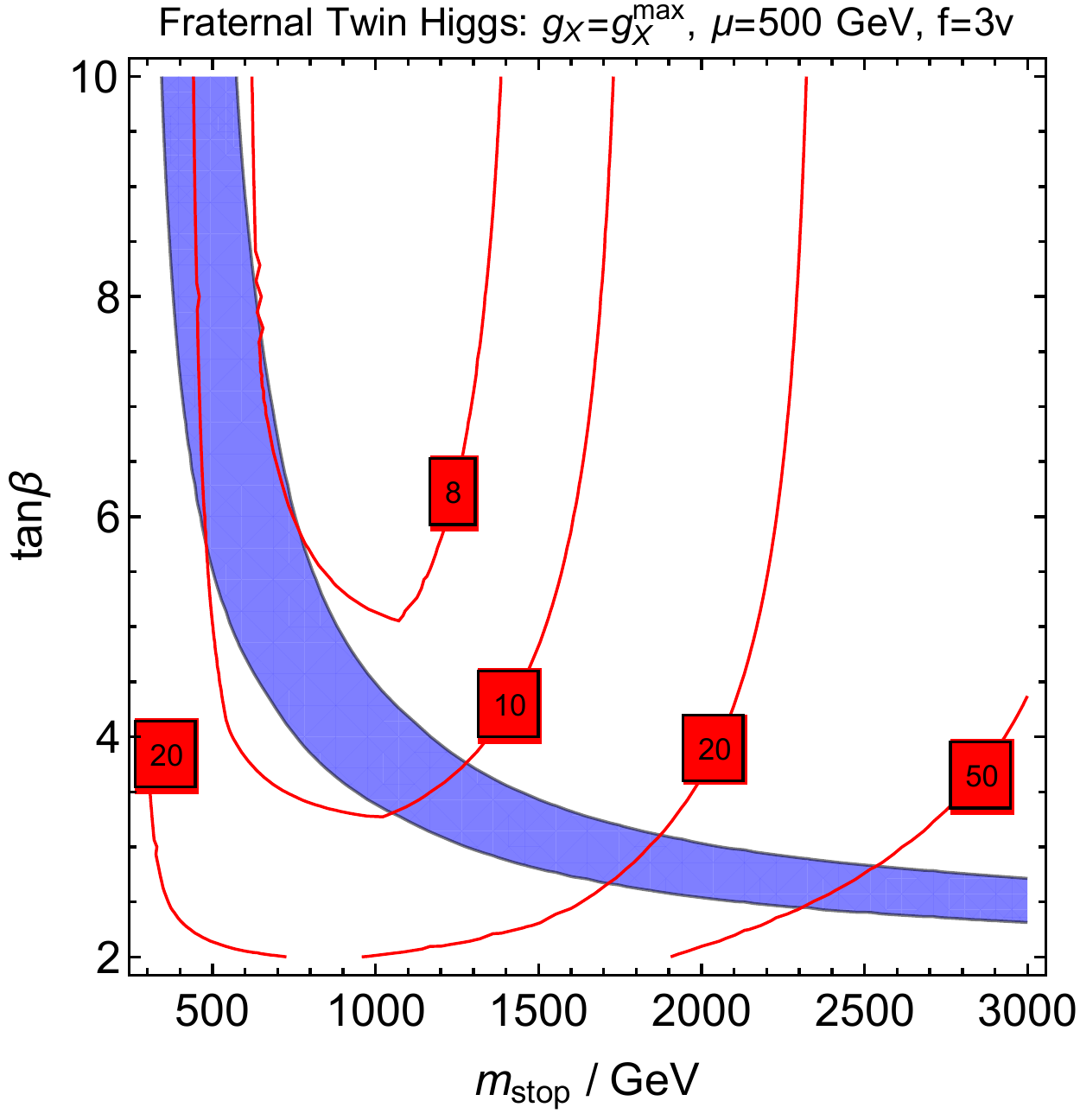}
\caption{Fine-tuning (red contours) in the $D$-term  Twin Higgs model for $f=3v$, $\mu=500$ GeV, $m_A=1$ TeV and $M_3=2$ TeV  assuming the messenger
scale $\Lambda=100 \Msusy$. In the left panels, where $\tan\beta=10$, the orange contours depict the value of the $SU(4)$ preserving quartic coupling
and in the green regions the Landau pole of the $U(1)_X$ gauge coupling constant is below $\Lambda$. In the right panels, at each point of the plane
$\Msusy$-$\tan\beta$, $g_X$ is fixed to the maximal value that allows the messenger scale to be below the Landau pole. In the blue region the Higgs
mass is in agreement with the measured value and several blue contours of the Higgs mass are also shown. In the top (bottom) panels mirror (fraternal)
Twin Higgs model is assumed.
}
\label{fig:Dterm}
\end{figure}

In order to minimize fine-tuning we demand that the fine-tuning due to the threshold correction of eq.~\eqref{deltamHu_X} does not exceed
the fine-tuning due to SUSY particles (dominated by stops, higgsino and gluino)
\begin{equation}
\label{finetune_X}
\Delta_{f,X} \equiv \frac{ \left(\delta m_{H_u}^2\right)_{X}}{2 \lambda f^2} <
\Delta_{f,{\rm SUSY}}.
\end{equation}
For a given value of $g_X$ and $\Msusy$, as well as gluino and higgsino masses, the
fine-tuning is minimized for the smallest value of $m_X$ allowed by experiments and $\epsilon$ chosen such that the inequality in
eq.~(\ref{finetune_X}) is saturated. The optimal value of $\epsilon^2$ is $\mathcal{O}(0.1)$ among most of the parameter space.%
\footnote{For large stop masses $\epsilon^2$ is much smaller than $0.1$ but the tuning would not be significantly different if
$\epsilon=0.1$ is taken instead because for such value of $\epsilon$ the SU(4) invariant coupling is already close to maximal,
see~eq.~\eqref{eq:lambdaD}. } 
Such values of $\epsilon$ do not require unusual hierarchies in underlying model, see Appendix for details where we present an explicit model generating
a non-decoupling $D$-term potential and show that other possible sources of tuning are less important than those included in our numerical analysis. 
We show the resulting contours of fine-tuning in the plane $\Msusy$-$g_X$ in
the left panels of fig.~\ref{fig:Dterm}.
We find that the effect of the $U(1)_X$ gauge coupling constant on the RG running of the yukawa coupling is important, as it reduces the top yukawa
coupling at higher energy scales.
As a result fine-tuning tends to be better for larger $g_X$ despite of larger $m_X$. This is another advantage with respect to the $F$-term model,
where the singlet effects tend to increase the top Yukawa coupling at higher energy scales.

The magnitude of tuning is different between the mirror and fraternal Twin Higgs models because the RG running of the $U(1)_X$
gauge coupling constant is faster in the latter case so for a given Landau pole scale $g_X$  must be smaller.
The top left panel of fig.~\ref{fig:Dterm} shows that in the mirror model with ${\rm tan}\beta=10$, $\Delta_v$ can be as small as about 8 for the stop
masses  up to about 1.2 TeV which corresponds to $\Delta_f\lesssim2$ i.e. essentially no fine-tuning in $f$.
Moderate tuning of the EW scale of 10~\% can be obtained for stops as heavy as about 1.4 TeV. In the fraternal model the same level
of fine-tuning as in the mirror model may be obtained for stops heavier by few hundred GeV, as seen from the bottom left panel of
fig.~\ref{fig:Dterm}.  For both mirror and fraternal models the tuning is dominated by the higgsino for the stop masses below about 1 TeV. This is
because we set $\mu=500$ GeV to evade the LHC constraints on sub-TeV sbottoms \cite{LHCsbottom}.  
Therefore, the current constraints on the  stop/sbottom masses
\cite{LHCstop}
do not introduce fine-tuning in $f$ in the $D$-term model. It may be even possible to have only moderate tuning of $\mathcal{O}(5\div10)\ \%$ even if
there is no sign of stops/sbottoms at the end of the high-luminosity run of the LHC. A useful reference point to compare is the non-twinned version of
the MSSM with non-decoupling $U(1)$ $D$-term in which tuning is worse by a factor $2\lambda/\lambda_{\rm SM}$. We see from fig.~\ref{fig:Dterm}
that in the $D$-term mirror (fraternal) Twin Higgs model the tuning may be smaller by a factor of about 3 (4).

It should be noted, however, that for $\tan\beta=10$ the Higgs mass constraint prefers rather light stops in the range between about 500 and 700 GeV
because heavier stops overshoot the experimental value of the Higgs mass. Such light stops (and sbottoms) are not in conflict with the current LHC results
 for the higgsino mass of 500 GeV that we use in our calculation of fine-tuning but this indicates that they may be discovered relatively
soon. One can imagine extensions of the present model in which there are negative contributions to the Higgs mass that may lift the stop masses
required to obtain the 125 GeV Higgs without altering tuning of the model. For example, one may consider an extension of the $D$-term Twin Higgs 
model by a singlet that couples in the superpotential only to the visible Higgs bosons in the NMSSM-like way i.e. $\lambda_S S H_u H_d$. Such
coupling generates a mixing between a singlet and the Higgs.  The mixing
effects between a relatively light singlet, say below 1 TeV, and the Higgs can provide a necessary reduction of the Higgs mass
without reduction of the $SU(4)$ preserving quartic term. \footnote{The impact of the Higgs-singlet mixing on the Higgs mass is similar as in the
NMSSM which was analysed e.g. in refs.~\cite{Hall:2011aa,Jeong:2012ma,Agashe:2012zq,Badziak:2013bda}.  } We leave a detailed analysis of such a model
for future work but we expect that the presence
of such singlet does not introduce
additional tuning. For example, the reduction of the Higgs mass by 15 GeV for the singlet mass of 500 GeV and $\mu=500$ GeV, as used in
our numerical example,  requires $\lambda_S\approx0.3$ which results in a subdominant correction to the soft Higgs mass from the singlet.

Alternatively, one can obtain the correct Higgs mass for heavier stops without extending the $D$-term model by reducing $\tan\beta$, as seen from the
right panels of fig.~\ref{fig:Dterm}. Smaller $\tan\beta$ reduces the tree-level Higgs mass but it also reduces the $SU(4)$ preserving quartic
couplings, so for given stop masses the tuning gets slightly worse for smaller $\tan\beta$. In order to increase the stop masses up to 1 (2) TeV
consistently with the 125 GeV Higgs mass, 
$\tan\beta$ must be reduced to about 4 (3) which increases tuning by only about 20 (50)~\% as compared to the $\tan\beta=10$ case. In consequence, 
even after taking the Higgs mass constraint
into account tuning better than 10~\% can be obtained for the stop masses up to about 1.2 (1.3) TeV in the mirror (fraternal) case.

\begin{figure}
\centering
\includegraphics[width=.49\textwidth]{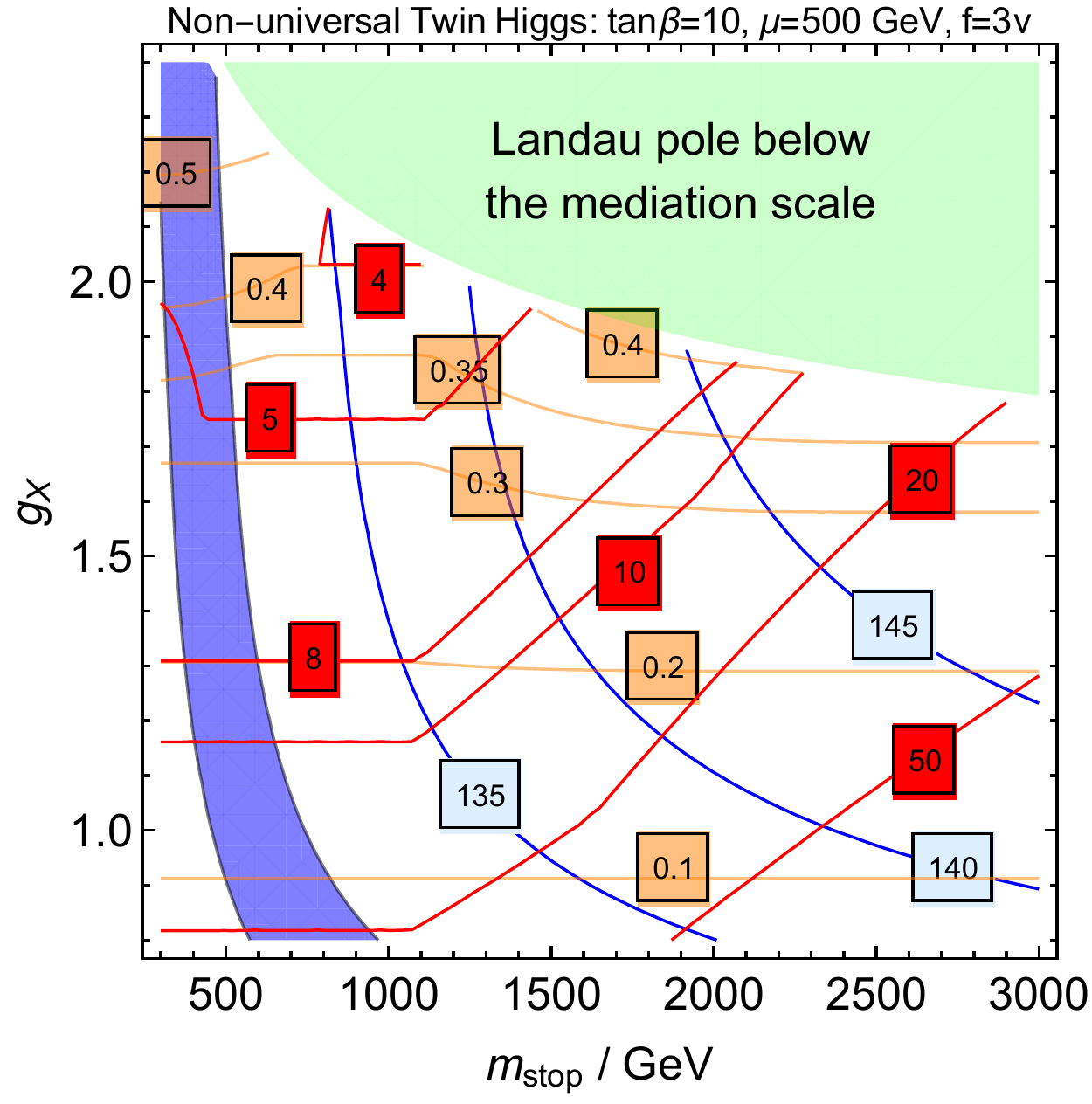}
\includegraphics[width=.49\textwidth]{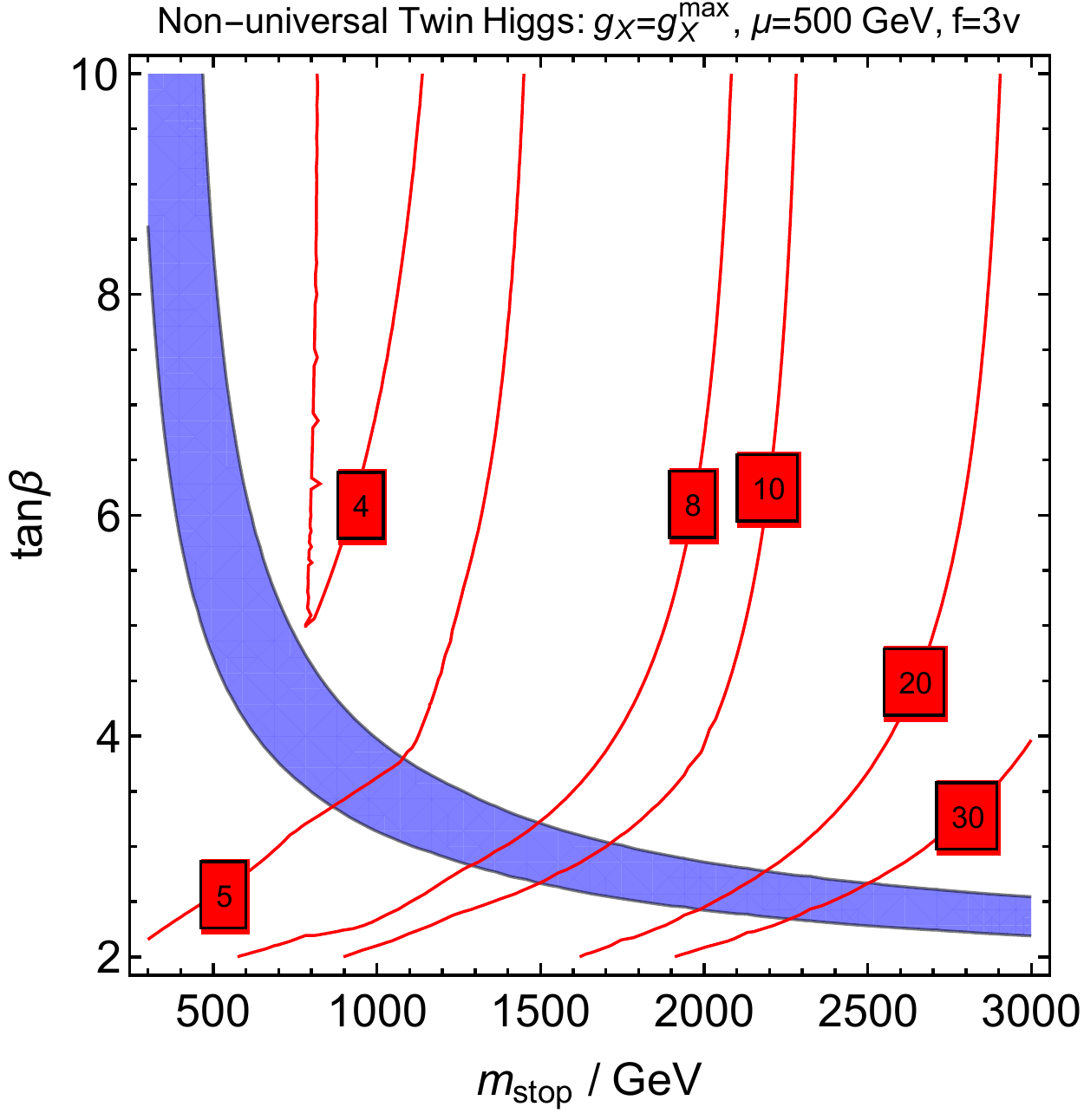}
\caption{The same as in fig.~\ref{fig:Dterm} but for the flavor non-universal $D$-term model with $M_X=5$~TeV. Similarly as in the case of
fig.~\ref{fig:Dterm}, $\epsilon$ is determined by the requirement that the condition in eq.~\eqref{finetune_X} is saturated. 
}
\label{fig:Dterm_only3rd}
\end{figure}

Let us also discuss the flavor non-universal model in which the first two generations of SM and mirror fermions are neutral under $U(1)_X$.
In such a case the $X$ gauge boson production at colliders is strongly suppressed and
there is no relevant constraint on $m_X$. In addition, below the scale of $U(1)_X$ charged fermions masses (which we reasonably assume to be at least two
orders of magnitude above $\Msusy$), the RG running of $g_X$ is slower. As a result larger values of $g_X$ may be obtained and the tuning may be
further
relaxed. We see from the left panel of fig.~\ref{fig:Dterm_only3rd} that in this case for $\tan\beta=10$ the tuning better than 25~\% can be obtained
even for the stop masses around 1 TeV, while for 2 TeV stops the tuning may be better than 10~\%. Since $\lambda$ can be as large as about 0.5,  the
tuning may be relaxed by a factor of  about 8
with respect to a non-twinned model with a non-decoupling $D$-term. Similarly as in the previous cases, the Higgs mass is overshot unless some
negative contribution to the Higgs mass is introduced. Nevertheless, even if no such negative correction is introduced the model may have better than
10~\% tuning for stop masses up to about 1.7 TeV consistently with the measured Higgs mass by taking smaller values of $\tan\beta$, as
demonstrated in the right panel  of fig.~\ref{fig:Dterm_only3rd}.

\subsection{Heavy Higgs spectrum}
\label{sec:spectrum}
Let us also discuss the heavy Higgs spectrum.
In the limit of $\lambda \gg \Delta \lambda$, $f\gg v$, the mass spectrum of the heavy Higgs in the $D$-term model is given by
\begin{align}
m_H^2 = &~m_A^2  =  m_{A'}^2 = m_{H^{'\pm}}^2 = 2 \mu^2 + m_{H_u}^2 + m_{H_d}^2 , \nonumber \\
m_{H'}^2 =&~\frac{1}{2}\left[ m_{A}^2 + \frac{g_X^2}{2} f^2 + \sqrt{ \left( m_{A}^2 - \frac{g_X^2}{2} f^2\right)^2 + 2 m_{A}^2 g_X^2 f^2 {\rm sin}^22\beta }  \right] \nonumber \\
\simeq &~ m_{A}^2 + \frac{g_X^2}{2} f^2 {\rm sin}^2(2\beta),\nonumber \\
m_{h'}^2 =&~\frac{1}{2}\left[ m_{A}^2 + \frac{g_X^2}{2} f^2 - \sqrt{ \left( m_{A}^2 - \frac{g_X^2}{2} f^2\right)^2 + 2 m_{A}^2 g_X^2 f^2 {\rm sin}^22\beta }  \right] \nonumber \\
\simeq &~  \frac{g_X^2}{2} f^2 {\rm cos}^2(2\beta)
\end{align}
In the second line of the expression of $m_{H'}^2$ and $m_{h'}^2$ we assume $g_X^2 f^2 \ll m_{A}^2$ and/or large ${\rm tan}\beta$. The Higgs mass
spectrum in the $F$-term model is derived in~refs.~\cite{Craig:2013fga,Katz:2016wtw}. We show the comparison of the spectra in the $D$ and $F$ term
models in Table~\ref{tab:Hspectrum}  assuming the decoupling limit, $\lambda^2 f^2 \ll m_{A}^2$.

\begin{table}[t]
\begin{center}
\begin{tabular}{|c||c|c|c|}
\hline
 & $h' \sim H_u'$ & $H\sim {\rm Re}(H_d^0)$ &   $H'\sim {\rm Re}(H_d^{'0})$  \\
\hline \hline
 $D$ & $4 \lambda f^2$ & $m_{A}^2$ & $m_{A}^2 + 4\lambda f^2 {\rm tan}^2 2\beta$  \\
 \hline
 $F$ & $4 \lambda f^2$ & $m_A^2$ & $m_A^2 + \frac{4 \lambda f^2}{{\rm tan}^2(2\beta)}$  \\ \hline
\end{tabular}
\begin{tabular}{|c||c|c|c|c|}
\hline
 &  $A \sim {\rm Im}(H_d^0)$ &  $A' \sim {\rm Im}(H_d^{'0})$ & $H^\pm \sim H_d^\pm$ & $H^{'\pm}  \sim H_d^{'\pm}$ \\
\hline \hline
 $D$  & $m_{A}^2$ & $m_{A}^2$ & $m_{A}^2$ & $m_{A}^2$ \\
 \hline
 $F$ &  $m_A^2$ & $m_A^2 + \frac{4 \lambda f^2}{{\rm sin}^2(2\beta)}$ & $m_A^2$ & $m_A^2$ \\ \hline
\end{tabular}
\caption{The mass spectrum of non-SM like Higgs bosons in the $F$-term and $D$-term Twin Higgs models in the $U(4)$ symmetric and $f\gg v$ limit.}
\label{tab:Hspectrum}
\end{center}
\end{table}%

In the $F$-term model there is  a potentially large radiative correction to $m_{H_u}^2$ from heavy Higgs doublets which is proportional to
$\lambda_S^2$. In order not to make fine-tuning even worse one expects the masses of $H/A$ to be rather small, potentially within the reach of the
LHC. In the $D$-term model naturalness also generically prefers light MSSM-like Higgs bosons and their twins. This is because $m_{H_d}^2$ gives
contribution to RGEs of $m_{H_u}^2$ through the trace of the soft mass squared weighted by $U(1)_X$ charges:
\begin{align}
            16 \pi^2 \frac{{\rm d} m_{H_u}^2}{{\rm dln} \mu}\supset g_X^2S  \,, \\
                   S={\rm Tr}(q_{X,i}m_i^2) \supset 2(m_{H_u}^2 - m_{H_d}^2) \,.
\end{align}
In the leading-log approximation the correction is
\begin{equation}
 \delta m_{H_u}^2\approx -\frac{g_X^2}{8\pi^2}m_{H_d}^2\ln\left(\frac{\Lambda}{m_X}\right) \,.
\end{equation}
Note that this correction differs from that from stops by a factor of
\begin{equation}
 -\frac{g_X^2m_{H_d}^2\ln(100\Msusy/m_X)}{6y_t^2 \Msusy^2 \ln(100)} \,.
\end{equation}
This implies that for $g_X\sim2$, $m_{H_d}^2\lesssim\Msusy^2$ is required to avoid increasing tuning. Therefore, MSSM-like Higgs bosons and
their
twins are expected to have their masses below 1 TeV since $m_A^2\approx m_{H_d}^2+ \mu^2 - 2\lambda f^2$. 
It is possible to avoid this conclusion if the trace $S$ vanishes while $|m_{H_u}^2| \ll m_{H_d}^2$ for some reason,
or $H_d$ is neutral under the $U(1)_X$ gauge symmetry; see ref.~\cite{Craig:2012bs} for a model with the latter property.
In such a case the quartic coupling from the $D$-term of the $U(1)_X$ gauge symmetry approximately preserves the $SU(4)$ symmetry for large $\tan
\beta$.

Let us take a closer look at the possible signatures of the Higgs sector of the $D$-term model at the LHC and compare it with
that of the $F$-term model studied in detail in ref.~\cite{Katz:2016wtw}.

\subsubsection*{Mirror Higgs $h'$}
The phenomenology of the mirror Higgs $h'$ is similar to that of generic Twin Higgs models \cite{Barbieri:2005ri,Buttazzo:2015bka} up to
non-decoupling effects due to heavy Higgs doublets, see the discussion on $H'$ below. The mass formula
for the mirror Higgs $h'$ is the same in both models but for a given $f$ in the $D$-term model one expects it to be somewhat
heavier due to a larger
achievable value of the $SU(4)$ preserving quartic coupling. In the least-tuned region with $f\lesssim 3v$ and $\lambda\sim0.3\div0.5$  the mirror
Higgs mass in the
$D$-term model is in the range of 500 to 700 GeV.

\subsubsection*{Mirror CP-even heavy Higgs $H'$}
The main difference between the $D$-term and $F$-term models is that in the former one there is large  mass degeneracy between $H$ and $H'$, see
table~\ref{tab:Hspectrum}. In the limit $m_A^2 \gg 4\lambda f^2$, this generically implies that
$H'$ has large $H_d$ component, unless $\tan\beta$ is very large. However, for sub-TeV masses of $H'$ which is phenomenologically interesting and is
also preferred to avoid the fine-tuning,
$H_u$ component of $H'$ is the most important for its production at the LHC as well as decay into visible final states. In contrast to the $F$-term model, in
the $D$-term model it is possible that mass of $H'$ is close to the mass of $h'$ which happens for $m_A^2 \sim
4\lambda f^2$ and results in a large $H_u$ component of $H'$, not much smaller than $v/f$. In such a case all three heavy CP-even states significantly mix
with each other. Therefore, it may be possible to observe three resonances, $h'$, $H$ and $H'$, with $m_H$ and $m_{H'}$ within few tens of GeV from
each other and close to $m_{h'}$. Detailed phenomenology depends on $\lambda$, $\tan\beta$, $m_A$ and $f$ and we leave a dedicated study of this issue
for future works.

\subsubsection*{Mirror CP-odd heavy Higgs $A'$}
In the $D$-term model, the mirror CP-odd heavy Higgs $A'$ does not mix with the MSSM Higgs bosons,
since the scalar potentials of the SM Higgs sector and the mirror Higgs sector have independent CP symmetries, $H_{u,d}\rightarrow H_{u,d}^\dag$ and
$H_{u,d}'\rightarrow H_{u,d}^{'\dag}$.
This should be contrasted with the $F$-term model, where $A'$ mixes with $A$ through the quartic coupling generated by the $F$ term of the singlet.
For this reason we do not discuss the phenomenology of $A'$, but we note that there may be a mixing if one introduce additional interactions, e.g.~a
superpotential coupling with a singlet field like in the $F$-term model.

\subsubsection*{Mirror charged Higgs $H^{'\pm}$}
$H^{'\pm}$ does not mix with any SM particles as long as the $U(1)_{\rm EM}'$ symmetry is unbroken.

\section{Conclusions}
\label{sec:concl}

We proposed a new SUSY UV completion of the Twin Higgs model in which the $SU(4)$ invariant quartic term $\lambda$ is provided by a $D$-term potential
from a new $U(1)_X$ gauge symmetry. In this setup $\lambda$ is maximized at large $\tan\beta$, which makes it possible to accommodate the
125 GeV Higgs mass simultaneously with the value of  $\lambda$ as large as about 0.5, and to greatly relax tuning of the EW scale. We
found
that the current LHC constraints can be satisfied with tuning better than 20~\%, while 2 TeV stops, which would be beyond the reach of the
LHC, may imply only moderate tuning of about 10~\%. This should be compared with the model in which $\lambda$ originates from an $F$-term of a new
singlet that results in the tuning of 2~\% at best.

We also discussed implications of the measured Higgs mass on the stop mass scale in a general SUSY Twin Higgs model in which the only source of the tree level $SU(4)$ breaking quartic term is the EW $D$-term potential. In particular, we found that in the large $\tan\beta$ limit of such models the Higgs mass
is larger than the measured value unless the stops are lighter than about 500 GeV. This region is already quite constrained by the LHC
experiments and not too light LSP is required to evade the bounds. These findings are especially interesting in the context of the $D$-term Twin
Higgs model proposed in this article, in which the least tuned region has large $\tan\beta$ and is expected to be covered by the LHC
stop/sbottom searches in the near future. Nevertheless, light stops are not a firm prediction of this model since the 125 GeV Higgs mass can be
obtained for heavier stops with smaller $\tan\beta$. For example, 1 (2) TeV stops require $\tan\beta$ of about 4 (3) which
makes tuning worse only by about 30 (60)~\% as compared to the large $\tan\beta$ result. Alternatively, the correct Higgs mass for heavy stops and
large $\tan\beta$ can be obtained  by introducing a negative contribution to the Higgs mass which can originate, for example, from a
mixing with a non-decoupled singlet.

If all the sparticles are pushed above the LHC reach, the only way to probe Twin Higgs models is via searches for additional Higgs bosons.  
We identified several differences in the heavy
Higgs spectrum  between the $D$-term and $F$-term model, which may help to distinguish them if several new scalars are
found in the future. 

\section*{Acknowledgments}
The authors would like to thank Lawrence Hall, Simon Knapen and Stefan Pokorski for useful discussions. 
This work has been partially supported by National Science Centre, Poland, under research grant DEC-2014/15/B/ST2/02157, by
the Office of High Energy Physics of the U.S. Department of Energy
under Contract DE-AC02-05CH11231, and by the National Science Foundation
under grant PHY-1316783. MB acknowledges support from the
Polish 
Ministry of Science and Higher Education through its programme Mobility Plus (decision no.\ 1266/MOB/IV/2015/0).

\appendix

\section{D-term potential and correction to Higgs soft masses}

In this appendix we discuss a model to break the $U(1)_X$ gauge symmetry, and the resulting $D$ term potential of the Higgs doublets as well the soft masses of them.
We introduce chiral multiplets $Z$, $P$ and $\bar{P}$, whose $U(1)$ charges are $0$, $+q$ and $-q$, respectively, and the following superpotential,
\begin{align}
\label{eq:X breaking}
W= \kappa Z (P \bar{P} -M^2),
\end{align}
where $\kappa$ and $M$ are constants. We assume that soft masses of $P$ and $\bar{P}$ are the same,
\begin{align}
V_{\rm soft} = m_P^2 \left( |P|^2 + |\bar{P}|^2 \right).
\end{align}
Otherwise, the asymmetric VEVs of $P$ and $\bar{P}$ give large soft masses to the Higgs doublets through the $D$-term potential.
The VEVs of $P$ and $\bar{P}$ are given by
\begin{align}
\vev{P} = \vev{\bar{P}} = \sqrt{M^2 - m_P^2 / \kappa^2} \equiv v_P.
\end{align}
The mass of the $U(1)_X$ gauge boson is given by
\begin{align}
m_X^2= 4 g_X^2 q^2 v_P^2.
\end{align}

In the SUSY limit, $m_P^2 \ll \kappa^2 M^2$, the $D$ term potential of the $U(1)_X$ charged particles vanishes after integrating out $P$ and $\bar{P}$.
In fact, after integrating out  the scalar components of $P$ and $\bar{P}$, we obtain the $D$ term potential of the Higgs doublets,
\begin{align}
\label{eq:effective D}
V_D = \frac{1}{8}g_X^2 \left( |H_u|^2 -  |H_d|^2   \right)^2 \left( 1 - \frac{m_X^2 }{2 m_P^2 + m_X^2}\right).
\end{align}
It can be seen that $V_D$ vanishes for $m_P^2=0$. From the above we determine the value of $\epsilon^2$ introduced in eq.~\eqref{eq:VU1X}:
\begin{equation}
 \epsilon^2= \frac{m_X^2 }{2 m_P^2 + m_X^2} \,.
\end{equation}
We see that $\epsilon\sim\mathcal{O}(0.1)$ does not require $m_P$ much larger than $m_X$.

Although the RG running of the Higgs doublets from $P$ and $\bar{P}$ vanishes due to the identical soft masses for $P$ and $\bar{P}$,
the threshold correction around the $U(1)_X$ symmetry breaking scale necessarily gives the correction to the Higgs doublets. At the one-loop level, we find
\begin{align}
\label{eq:threshold correction}
\delta m_{H_u}^2 = \frac{g_X^2 m_X^2}{64\pi^2} {\rm ln}\frac{2 m_P^2 + m_X^2}{m_X^2} = \frac{g_X^2 m_X^2}{64\pi^2} {\rm ln}\left(\epsilon^{-2}\right).
\end{align}
Here we assume that the SUSY breaking contribution to the gaugino mass and the soft masses of Higgs doublets are negligible.
The results in Eqs.~(\ref{eq:effective D}) and (\ref{eq:threshold correction}) are consistent with the model-independent discussion in
ref.~\cite{Cheung:2012zq}.

Let us comment on the possible sources of the fine-tuning in addition to $\Delta_f$ due to the threshold correction.
In the above analysis we assume that the soft masses of $P$ and $\bar{P}$ are identical.
This can be guaranteed by a C symmetry $P\leftrightarrow \bar{P}$ in the sector which generates the soft masses.
The symmetry is explicitly broken by the SM $SU(2)\times U(1)$ gauge interaction and the yukawa interactions.
Once the soft masses of $P$ and $\bar{P}$ are different with each other, there are extra contributions to $m_{H_u}^2$, so the tuning $\Delta_f$
may become worse.
Among the source of the breaking of the symmetry $P\leftrightarrow \bar{P}$, the top yukawa coupling is the largest, and we expect that the following
magnitude of the soft mass difference is unavoidable,
\begin{align}
\Delta m^2 \equiv  m_P^2 -m_{\bar{P}}^2 \sim q \frac{g_X^2}{16\pi^2} \frac{y_t^2}{16\pi^2}m_{\rm stop}^2.
\end{align}
This results in the asymmetric VEVs of $P$ and $\bar{P}$,
\begin{align}
\vev{P}^2 - \vev{\bar{P}}^2 \simeq  -\Delta m^2 \frac{- m_P^2 + \kappa^2 M^2 }{2 g_X^2 \kappa^2 q^2 M^2 - 2 g_X^2 q^2 m_P^2 + \kappa^2 m_P^2} = - \frac{\epsilon}{2 g_X^2 q^2} \Delta m^2,
\end{align}
which gives the soft mass of $H_u$ through the $D$ term potential of the $U(1)_X$ gauge interaction,
\begin{align}
\label{eq:tree mHu}
\Delta m_{H_u}^2|_{D\text{-term}} = \frac{1}{2} g_X^2 q \left( \vev{P}^2 - \vev{\bar{P}}^2 \right) \simeq - \frac{\epsilon }{4 q} \Delta m^2 \simeq -
\frac{\epsilon}{4} \frac{g_X^2}{16\pi^2} \frac{y_t^2}{16\pi^2}m_{\rm stop}^2
\end{align}
The mass difference of the soft mass of $P$ and $\bar{P}$ also generates the soft mass of $H_u$ through the quantum correction via the $U(1)_X$ gauge
interaction,
\begin{align}
\label{eq:loop mHu}
\Delta m_{H_u}^2|_{\rm loop} \sim \frac{q g_X^2}{16\pi^2} \Delta m^2 \simeq q^2 \left( \frac{g_X^2}{16\pi^2}  \right)^2 \frac{y_t^2}{16\pi^2}m_{\rm stop}^2.
\end{align}
The corrections in eqs.~(\ref{eq:tree mHu}) and (\ref{eq:loop mHu}) are smaller than the usual one-loop correction to $m_{H_u}^2$ from the top yukawa
interaction by extra loop factors, 
and do not affect the fine-tuning $\Delta_f$.
This may be invalidated if the Landau pole scale of the $U(1)_X$ gauge interaction is close to the mediation scale of the SUSY breaking.
In this case, $\Delta m_{H_u}^2$ is expected be as large as the one in eqs.~(\ref{eq:tree mHu}) and (\ref{eq:loop mHu}) with $g_X^2/16\pi^2 = O(1)$.
The correction does not involve the logarithmic enhancement as $g_X$ becomes smaller for smaller energy scales.
We again expect that the correction to $m_{H_u}^2$ does not exceed the usual one-loop correction via the top yukawa interaction.

The parameter $\epsilon$ is given by the model parameters $M$, $m_P$, $\kappa$, $q$, $g_X$ as
\begin{align}
\epsilon^2 \simeq 2q^2 g_X^2 \left( \frac{M^2}{m_P^2} - \frac{1}{\kappa^2} \right),
\end{align}
where we assume $\epsilon \ll 1$.
Since the gauge coupling $g_X$ is large, in order to obtain $\epsilon^2={\cal O}(0.1)$ naturally,
a small $q$, say $1/6$, and/or a large $\kappa$ is required.
The latter option requires that the superpotential in eq.~(\ref{eq:X breaking}) is UV completed, e.g.~by the deformed moduli
constraint~\cite{Seiberg:1994bz}.
If neither $q$ is small nor $\kappa$ is large one needs some tuning between $M^2 / m_P^2 $ and $1/\kappa^2$.

\end{document}